\documentclass[journal]{IEEEtran}
\pdfoutput=1
\ifCLASSINFOpdf
\else
\fi

\usepackage{cite}
\usepackage{booktabs}
\usepackage{multirow}
\usepackage{amsmath,amssymb,amsfonts,amsthm}
\usepackage{algorithm}
\usepackage{algorithmic}
\usepackage{textcomp}
\usepackage{url}
\usepackage{bbm}
\usepackage{enumerate}
\usepackage{xcolor}
\usepackage{graphicx}
\usepackage{graphics}
\usepackage{subfigure}
\usepackage{pifont}

\ifodd 1
\newcommand{\com}[1]{\textbf{\color{red} (COMMENT: #1)}} 
\newcommand{\comg}[1]{\textbf{\color{green} (COMMENT: #1)}}
\newcommand{\response}[1]{\textbf{\color{magenta} (RESPONSE: #1)}} 
\else

\newcommand{\com}[1]{}
\newcommand{\comg}[1]{}
\newcommand{\response}[1]{}
\fi

\begin{document}

\title{A3D: Adaptive, Accurate, and Autonomous Navigation for Edge-Assisted Drones}

\author{Liekang~Zeng,
        Haowei~Chen,
        Daipeng~Feng,
        Xiaoxi~Zhang,
        and~Xu~Chen
\thanks{
The authors are with the School of Computer Science and Engineering, Sun Yat-sen University, Guangzhou, Guangdong, 510006 China (e-mail: \{zenglk3, chenhw26, fengdp3\}@mail2.sysu.edu.cn, \{zhangxx89, chenxu35\}@mail.sysu.edu.cn).
}
}

\maketitle

\begin{abstract}
Accurate navigation is of paramount importance to ensure flight safety and efficiency for autonomous drones.
Recent research starts to use Deep Neural Networks (DNN) to enhance drone navigation given their remarkable predictive capability for visual perception.
However, existing solutions either run DNN inference tasks on drones in situ, impeded by the limited onboard resource, or offload the computation to external servers which may incur large network latency. Few works consider jointly optimizing the offloading decisions along with image transmission configurations and adapting them on the fly. 
In this paper, we propose A3D, an edge server assisted drone navigation framework that can dynamically adjust task execution location, input resolution, and image compression ratio in order to achieve low inference latency, high prediction accuracy, and long flight distances.
Specifically, we first augment state-of-the-art convolutional neural networks for drone navigation and define a novel metric called Quality of Navigation as our optimization objective which can effectively capture the above goals. 
We then design a deep reinforcement learning (DRL) based neural scheduler at the drone side for which an information encoder is devised to reshape the state features and thus improve its learning ability. 
To further support simultaneous multi-drone serving, we extend the edge server design by developing a network-aware resource allocation algorithm, which allows provisioning containerized resources aligned with drones' demand.
We finally implement a proof-of-concept prototype with realistic devices and validate its performance in a real-world campus scene, as well as a simulation environment for thorough evaluation upon AirSim.
Extensive experimental results show that A3D can reduce end-to-end latency by 28.06\% and extend the flight distance by up to 27.28\% compared with non-adaptive solutions.
\end{abstract}


\IEEEpeerreviewmaketitle

\section{Introduction}
Recent years have witnessed a growing deployment of autonomous drones in various real-world scenarios, such as search and rescue in natural disasters, smart agriculture, and smart cities \cite{mishra2020drone,vasisht2017farmbeats,alsamhi2019survey}.
While the advanced ability in image/video content perception and analytics has made Deep Learning (DL) techniques a \textit{de-facto} standard tool for visual applications \cite{bengio2017deep}, autonomous drones are becoming more intelligent and serviceable by carrying Deep Neural Networks (DNNs) for navigation guidance.
Specifically, in a typical DL-enabled flight, a DNN model accepts images captured by the drone's camera continuously, and exports a steering angle and a flying velocity to steer the control of aerofoils, and therefore reacts to the dynamic physical environments.

\begin{figure}[t]
    \centering
    \setlength{\abovecaptionskip}{0cm}
    \subfigure[Delayed navigation decision may lead the drone to a crash.]{
        \begin{minipage}[t]{0.45\linewidth}
        \centering
        \includegraphics[width=\linewidth]{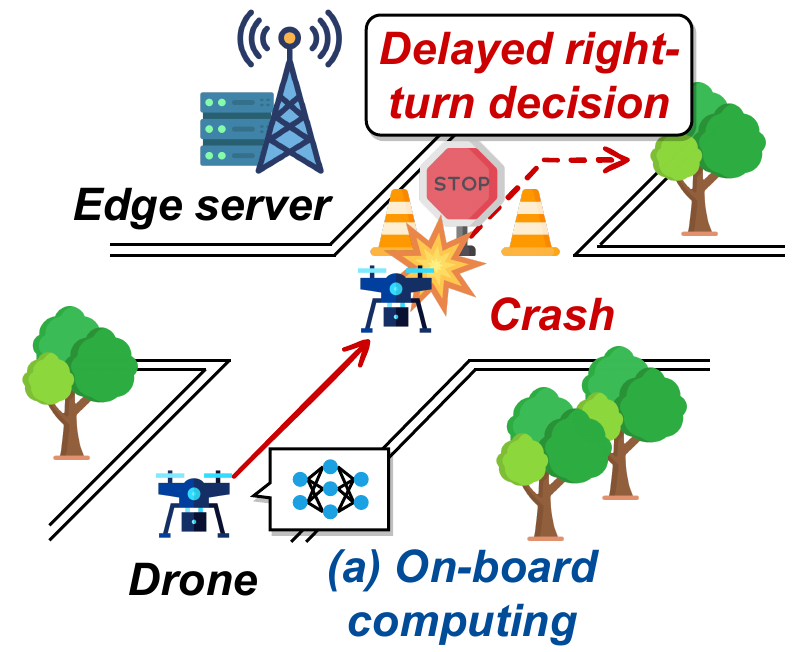}
        \label{fig:scenario_delayed}
        \vspace{-0.5cm}
        \end{minipage}
    }
    \
    \subfigure[Timely navigation decision steers the drone to a safe trajectory.]{
        \begin{minipage}[t]{0.45\linewidth}
        \centering
        \includegraphics[width=\linewidth]{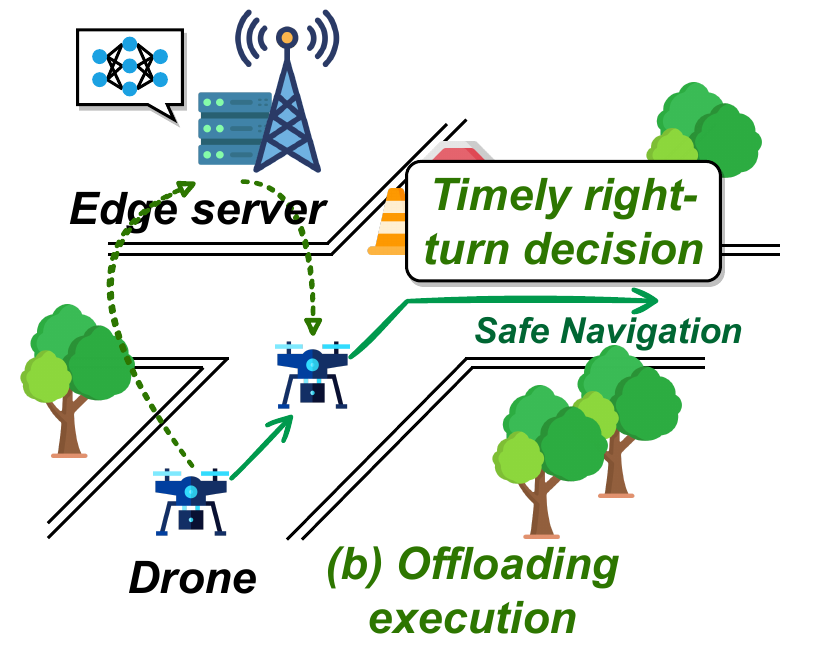}
        \label{fig:scenario_timely}
        \vspace{-0.5cm}
        \end{minipage}
    }
    \caption{Example scenario of an autonomous drone flying on a city road, where its expected navigation trajectory is to go straight and then turn right.  }
    \label{fig:scenario}
    \vspace{-0.5cm}
\end{figure}

While recent progress in DNN models has pushed navigation accuracy to an unprecedented altitude, deploying them in the physical world is up against a set of obstacles.
First, the climb of navigation accuracy comes with deeper, larger, and more sophisticated architectures, which in principle accompany heavier workloads and considerable energy consumption.
Running these resource-hungry DNN models onboard can thus dramatically reduce the available endurance time of power-limited drones.
Second, while existing DL models have achieved excellent navigation accuracy offline, the growing inference latency may conversely decline the navigation quality at runtime.
To illustrate that, Fig. \ref{fig:scenario} presents an example where a drone is self-flying on city roads.
With an image of a straight road captured at a starting location, the autonomous drone system may run an inference with its navigation model to continuously decide a route.
However, this inference task may take a prohibitively long time, resulting in a delayed right-turn decision at the crossroad (where a stop sign stands) and thus an unexpected crash and flight termination as shown in Fig. \ref{fig:scenario_delayed}.
As we measure in different routes (Sec. \ref{sec:hidden_dimension}), milliseconds of latency can significantly reduce the performance of navigation.
Worse still, lowering the exceedingly high inference latency is intractable due to the inherent conflict of computationally intensive DL workload and constrained computing capability of drones, hindering high-quality navigation in real deployment.

To overcome these problems simultaneously, in this paper, we leverage the emerging edge intelligence paradigm \cite{zhou2019edge} and propose A3D, a dynamic navigation framework that can adaptively collaborate drones with edge servers for high-quality autonomous flight.
As illustrated in Fig. \ref{fig:scenario_timely}, A3D eases the drone's burden by selectively migrating onboard workload to nearby edge servers, targeting reducing inference latency for accurate navigation decisions.
A3D's design goes beyond directly combining offloading with onboard computing for accelerating execution speed. Instead, it addresses the following three challenges.

{\em First}, while offloading execution embraces external computing resources for performance enhancement, it comes at a price of functional dependence on some environmental factors, such as network conditions and available edge resources, which can fluctuate during the flight. 
On this issue, many edge intelligent systems aim at optimizing accuracy under the constraint of latency \cite{galanopoulos2021automl,zhao2021edgeml}. 
However, in autonomous navigation, users prefer the drone's autonomy rather than solely latency or accuracy.
As we show in Sec. \ref{label:challenge}, latency and accuracy can affect autonomy in a complex relationship, and viewing them in a compartmentalized manner may lead to poor autonomy performance for navigation. 
Designing new metrics to better characterize the overall flight performance is called for.

{\em Second}, while a new performance metric combining latency and accuracy may not be hard to derive, mathematically optimizing the drone navigation process is hard, given that the environmental dynamics in the navigation routes and edge networks are uncertain and could have extreme variations. Besides, different controllable decision variables rooted in optimizing image transmission configurations and leveraging edge computing need to be solved simultaneously, enforcing the problem to be combinatorial, further hindering solving for the optimal solutions in real time. To address this, we adopt Deep Reinforcement Learning (DRL) to combat the uncertainty and learn the joint optimization through errors and trials.

{\em Third}, directly applying off-the-shelf DRL algorithms is insufficient for our scenario given that the observable states in the drone navigation environment construct a large search space and may contain indirect information that affects decision making. Therefore, enhanced state abstraction is needed to encode the raw states into better learnable features rather than directly feeding the observable ones into the DRL model. Moreover, the scheduler needs to be implemented in a lightweight manner so that the scheduling is viable given that the navigation inference already has potentially large latency which is why we enable task offloading in the first place.

To address the challenges, we make the following technical contributions.
\begin{itemize}
    \item We make a comprehensive investigation on edge-assisted navigation model inference for autonomous drones, revealing the complex nexus between inference latency and accuracy. To organically combine both metrics, we treat \textit{autonomous navigation as a service} and formally define a novel and comprehensive metric called \textit{Quality of Navigation} (QoN), to quantify the overall scheduling performance. By regarding each navigation decision inference as a service attempt and setting a threshold of prediction error, QoN essentially characterizes the success rate of navigation decision within a time window of flight so as to capture inference latency and accuracy simultaneously.
    
    \item We develop a DRL-based neural scheduler to learn the optimal scheduling policy for high-quality navigation with the goal of maximizing the overall QoN of the flight. An environmental information encoding module is additionally designed and incorporated as the front end into the scheduler. Serving as state abstraction enhancement, it enables the DRL agent to capture the dependency between different state features and their statistical characteristics in the dynamic environment, improving the learning efficiency.
    
    \item We propose A3D, a novel drone-edge synergetic framework for high-quality autonomous drone navigation with the assist of edge servers.
    A3D incorporates the neural scheduler at the drone side for adaptively scheduling the autonomous navigation tasks by simultaneously optimizing multiple configuration parameters and the task offloading decision. 
    At the edge server side, A3D applies a containerized environment to dynamically allocate edge resources for individual drones and serve navigation model inference queries.
    
\end{itemize}

\textbf{Supporting multiple drones.}
To enable A3D to support simultaneous multi-drone serving, we further extend our system design at the edge server with a dynamic resource allocation mechanism.
Specifically, we focus on improving the average QoN experienced by all connected drones through distributing proper edge resources for their corresponding serving containers (which host their navigation models).
From preliminary experiments, we observe that inference queries with heavier workload (e.g., input images with higher resolution) are more sensitive to resource replenishment, and the bandwidths between individual drones and the edge server can be utilized as an indicator to reflect how much they would like to offload their workload.
We therefore leverage an on-demand strategy and develop an intelligent resource allocation algorithm that is able to judiciously assign proper containerized resources at the edge server to drones for global performance boosting among them.

\textbf{Performance evaluation.}
We implement a proof-of-concept prototype of A3D using realistic testbeds and evaluate its performance in a campus route.
Experimental results demonstrate that A3D outperforms existing baselines by up to 21.97\% QoN improvement and achieves 1.18$\times$ flight distance extension.
To complement a thorough evaluation with more settings, we further implement a simulation environment upon the AirSim simulator and examine the performance for both single-drone and multi-drone serving.
Our simulation results show that A3D outperforms existing non-adaptive solutions, reducing inference latency by 28.06\% on average, and extending flight distance up to 27.28\%.
The multi-drone simulation on A3D against existing heuristics shows that our proposed resource allocation algorithm improves the average QoN by up to 13.6\%, while extending the average flight distance of drones for at most 42.07m.
In addition, A3D's neural scheduler (at the drone side) is particularly lightweight, introducing no more than 5ms running overhead to the navigation runtime, which can be applicable to other emerging DNN-driven autonomous navigation scenarios.

\textbf{Organization.}
The rest of this paper is organized as follows. 
Sec. \ref{sec:background} briefly reviews autonomous drone navigation and investigates the hidden optimization dimension for navigation performance.
Sec. \ref{sec:navigation_service} introduces the proposed QoN metric and discusses the configuration space and challenges of drone adaptability.
Sec. \ref{sec:overview} overviews the system design of A3D, and Sec. \ref{sec:neural_scheduler} and Sec. \ref{sec:support_multiple_drones} presents in detail the neural scheduler at the drone side and the resource allocator at the edge side, respectively.
Sec. \ref{sec:implementation} shows the implementation of our realistic prototype and simulation environment and Sec. \ref{sec:evaluation} provides the evaluation results.
Sec. \ref{sec:related_work} reviews the related works and Sec. \ref{sec:conclusion} concludes.

\begin{figure}[t]
    \centering
    \setlength{\abovecaptionskip}{-0.1cm}
    \setlength{\belowcaptionskip}{-0.5cm}
    \includegraphics[width=0.9\linewidth]{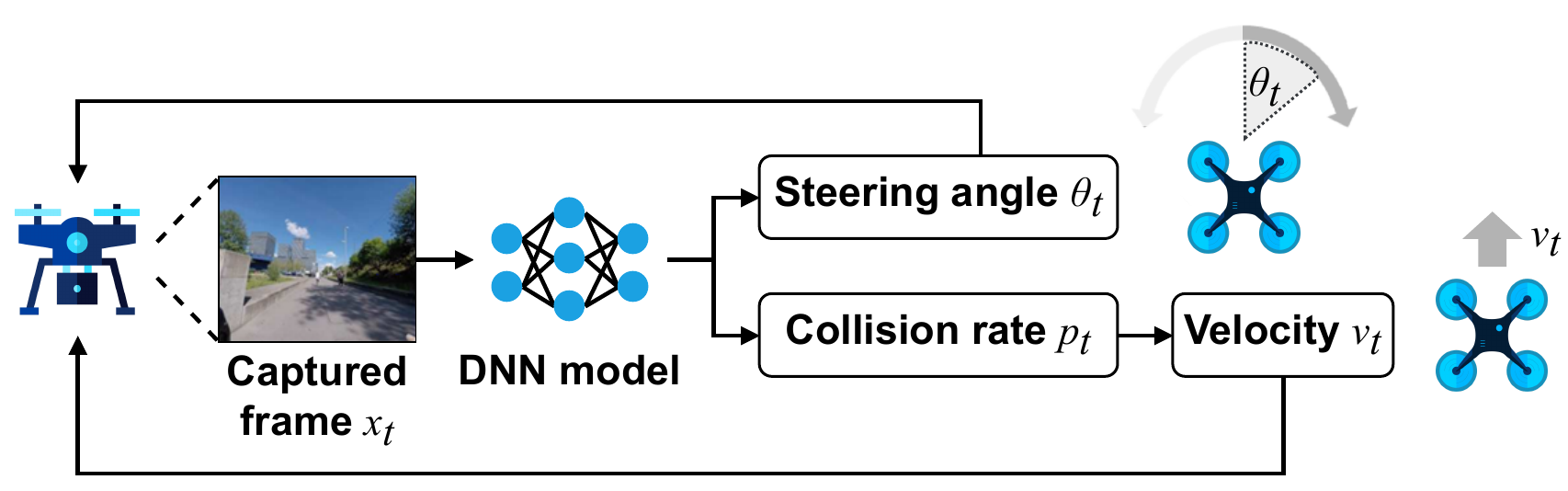}
    \caption{
    In each control loop, a drone captures an image $x_t$ and calls a DNN model to export a steering angle $\theta_t$ and a collision rate $p_t$, where the latter yields a velocity $v_t$.
    }
    \label{fig:control_loop}
    \vspace{-0.5cm}
\end{figure}

\begin{figure}[t]
    \centering
    \setlength{\abovecaptionskip}{-0.1cm}
    \setlength{\belowcaptionskip}{0.2cm}
    \includegraphics[width=0.9\linewidth]{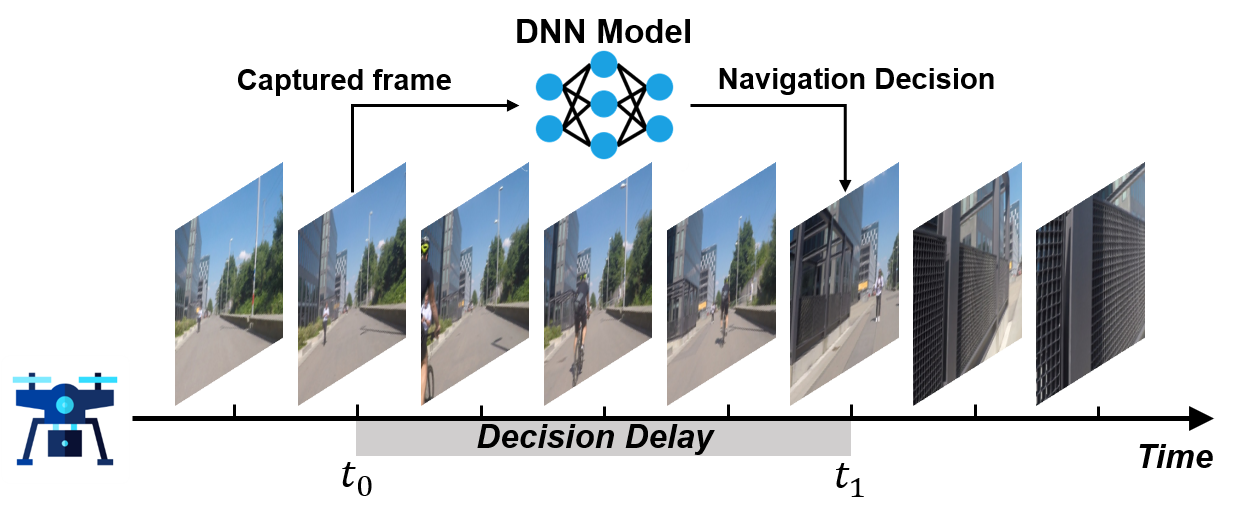}
    \caption{The prediction latency has been a hidden dimension that significantly impacts the optimization of safe and reliable navigation: the delay of navigation decision at time $t_0$ can yet lead the flying drone to a crash at time $t_1$.}
    \label{fig:navigation_delay}
    \vspace{-0.5cm}
\end{figure}

\section{Background and Motivation}
\label{sec:background}
\subsection{Autonomous Drone Navigation}

With the widely spread of unmanned applications, autonomous drones have been utilized in a variety of real-world scenarios ranging from path piloting \cite{smolyanskiy2017toward}, object detection \cite{zhu2018visdrone} to disaster rescue \cite{mishra2020drone}, \textit{etc.}
For example, autonomous drones have been employed in Amazon's delivery services \cite{AmazonPrimeAir} for on-demand unmanned product expresses.

At the core of these services, the self-sufficient navigation model is the fundamental component to enable autonomy.
In particular, we focus on navigating edge-assisted drones, where the vehicles are committed to flying through a legible route with the support of ground stations (i.e., edge servers). 
As their function heavily relies on accurate environmental perception, recent advances have applied powerful DNNs as navigation models to generate flying decisions\cite{jung2018perception}.
Fig. \ref{fig:control_loop} depicts a typical control loop of a DNN-driven navigation \cite{loquercio2018dronet}.
In each operating epoch, a drone scans the frontal landscape using its camera and passes the captured image $x_t$ to the DNN model for exporting a corresponding navigation decision.
Particularly, the decision comprises two parts.
One is the steering angle $\theta_t$, which is specified in the turning radian with respect to the current orientation and will be used to direct the turning obliquity of aerofoils at the next moment.
For instance, a right-turn command corresponds to a steering angle of $\pi/2$ ($90^{\circ}$), and a left-turn command is exactly $-\pi/2$ ($-90^{\circ}$).
The other is the collision rate $p_t$ which is used to generate the drone's forward velocity $v_t$ by linear transformation $v_t = v_{\text{max}}(1-p_t)$, where $v_{\text{max}}$ is the maximum drone speed.
With the DNN's output acting as feedback operating on the drone's flight module, the control procedure constructs a closed-loop and drives the navigation to react to physical world constantly.

\subsection{Hidden Dimensions in Accurate Navigation}
\label{sec:hidden_dimension}

One of the most critical requirements of autonomous navigation is safety, demanding a timely and accurate decision in dynamic environments. 
However, current work on CNN-based autonomous navigation ignores the impact of end-to-end latency on drone navigation performance.
As an example, Fig. \ref{fig:navigation_delay} illustrates an initial instant when the drone's camera captures an image as $t_0$ and the prospective moment when the navigation model outputs a decision with respect to that image as $t_1$.
Since the drone actually follows the command corresponding to input at $t_0$ rather than the real scene at $t_1$, the 
navigation decision can be \textit{expired}, which may lead to a yaw and even a crash.
We thus argue that latency is a hidden dimension in accurate autonomous navigation, which calls for joint optimization together with the accuracy metric to ensure an efficient and secure journey.

\begin{figure}[t]
    \centering
    \setlength{\abovecaptionskip}{0cm}
    \includegraphics[width=0.95\linewidth]{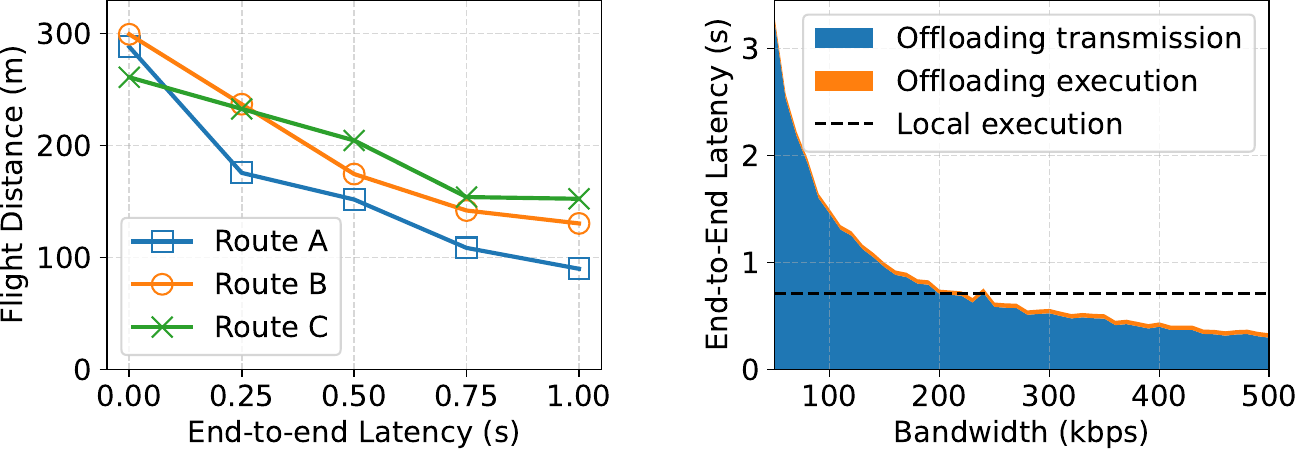}
    \caption{Left: As the end-to-end latency of navigation decisions increases, the achieved flight distance dramatically decreases. Right: End-to-end latency of offloading and local execution, where the offloading latency breaks down in communication and computation.}
    \label{fig:motivation_latency}
    \vspace{-0.5cm}
\end{figure}

The above analysis is further confirmed by quantitative measurements in AirSim simulator, with results shown in Fig. \ref{fig:motivation_latency}(left).
For each flight tour, we force the inference latency as a determined value and let the drone fly freely until it deviates from the expected route.
We record the flight distances upon their terminations, which is a common metric of navigation performance, and find that the achieved meters rapidly diminish as the navigation decision latency increases, across different types of routes.

Note that navigation accuracy can be oblivious of inference latency if the command from DNNs stays invariant, e.g., a constant ``go straight" signal in a long straight avenue.
However, real-world cases usually consist of many curves and crossroads, where any delay of decisions can dramatically decline navigation precision and the above conclusion holds.

\subsection{Limitations of Existing Solutions}

In the context of CNN-based navigation, existing works typically equip drones with powerful computing devices \cite{smolyanskiy2017toward,sanket2020evdodgenet} or assume stable network connectivity for drones to nearby servers \cite{kouris2018learning,loquercio2018dronet}, which is usually unavailable and unpractical in real-world scenes.
Towards lowering the delay of DNN inference, a number of works center on local computing and alleviate device's workload by employing smaller DNN architectures \cite{jiang2018chameleon} or augmenting devices with hardware accelerators \cite{tan2020fastva}.
However, neither of them enables a farther flight distance in that reducing navigation workload can decline the steering accuracy, and extending computing hardware increases power consumption to the tiny battery.
To utilize supplementary resources without additional onboard burden, another line of works resorts to offloading workload to nearby edge servers such as 5G MEC servers.
Nonetheless, their heavy dependence on wireless transmission makes them highly sensitive to network conditions, which are typically fluctuating and unstable due to drones' mobility.

We measure the costs of both ways by computing DroNet \cite{loquercio2018dronet}, a state-of-the-art drone navigation model, on a Jetson Nano (as the drone processor) and a desktop PC (as the edge server), adjusting the bandwidth between them. 
As reported in Fig. \ref{fig:motivation_latency}(right), the end-to-end latency of navigation decision is extremely high ($>$1.5s) when the bandwidth is very limited ($<$100kbps), and is even poorer than that of local execution on board (0.709s) though they all fail to meet real-time requirements.
Breaking down the costs of offloading we observe that the transmission stage dominates the entire performance, implying the exorbitant reliance on networking conditions.
Overall, we observe that both approaches have their advantages and limitations, presenting a prospective opportunity to combine them for real-time navigation.
This motivates us to design a joint optimization considering the nexus of latency and accuracy simultaneously, bridging the performance gap between local and offloading execution with adaptive decisions.

\begin{figure}[t]
    \centering
    \setlength{\abovecaptionskip}{0cm}
    \includegraphics[width=0.85\linewidth]{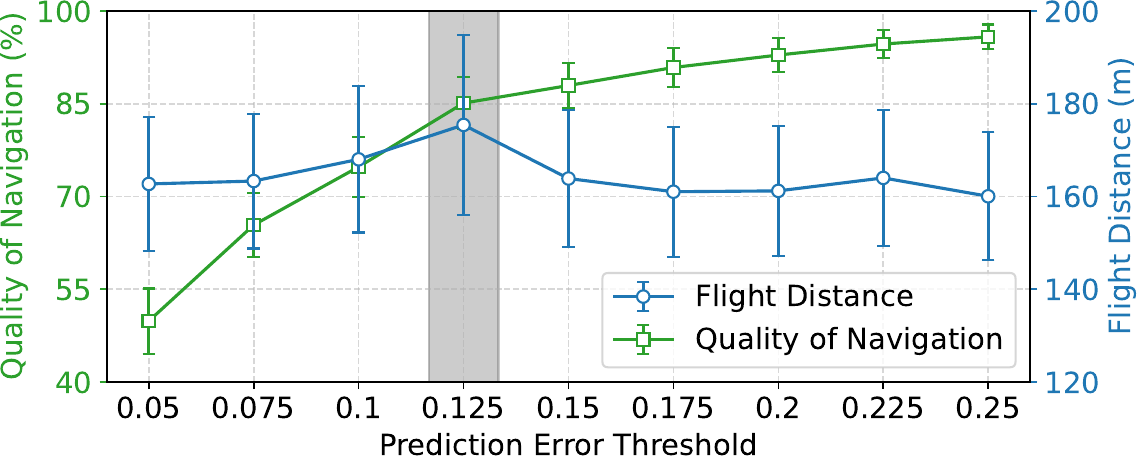}
    \caption{The Quality of Navigation and the flight distance of drones with different prediction error threshold $\varepsilon$, where we observe that setting $\varepsilon$ in $[0.11, 0.13]$ can achieve the optimal flight distance.}
    \label{fig:vary_threshold}
    \vspace{-0.5cm}
\end{figure}

\section{Adaptive Navigation as a Service}
\label{sec:navigation_service}

To characterize the performance of an accurate and adaptive autonomous drone flight in a more systematic way, we propose to treat \textit{adaptive navigation as a service} and study the navigation performance from a service perspective.
Specifically, we first formally define the quality of navigation and next discuss the design space and challenges of scheduling adaptability.

\subsection{Quality of Navigation Metric Design}
\label{sec:QoN}

Service Level Objective (SLO) is widely employed as a way of quantitative measurement of service performance.
For accurate navigation, we instantiate the SLO as a prediction error threshold $\varepsilon$ in steering angle deviation, indicating the user's tolerance in navigation precision.
Specifically, for any time $t$, given the model prediction on the turning angle as $\theta_{\text{pre}}^t$ based on the current input image captured at time $t$ and the ground truth as $\theta_{\text{gt}}^t$ based on the real-scene image at time $t+t_{\text{delay}}$ exactly, the navigation service should satisfy:
\begin{align}
    |\theta_{\text{pre}}^t - \theta_{\text{gt}}^t| \leq \varepsilon. \label{eq:error_threshold}
\end{align}
The unit of $\varepsilon$ is radian, which directly follows steering angle's unit. The smaller the $\varepsilon$ is, the stricter requirement the navigation precision expects.

Next we investigate how many times the navigation decision meets the SLO within a given time window $\tau$.
Particularly, each time a navigation decision is exported, we regard it as a service event towards the error threshold $\varepsilon$ and check a successful attempt if Eq. (\ref{eq:error_threshold}) holds and a defectiveness or else.
We can therefore interpret the \textit{Quality of Navigation} (QoN) by readily calculating the service success rate, i.e. the ratio between succeed times and total decision times, formally defined as:
\begin{align}
    \mathcal{Q} = \sum^{\tau}_{t=0} I(|\theta_{\text{pre}}^t - \theta_{\text{gt}}^t| \leq \varepsilon)/\tau,
    \label{eq:QoN}
\end{align}
where $I(\cdot)$ is an indicator function that returns 1 if the predicate feeds a true value. Note that collision rate is highly correlated with the turning angle since they are generated by the navigation model with the same input and backbone model, and hence collision rate is not considered to avoid redundancy in QoN calculation. 

In addition, the hyper-parameter $\varepsilon$ in QoN is scenario-dependent and can be tuned according to some more intuitive metrics (e.g., flight distance) in practice.
In general, $\varepsilon$ should not be set too large or too small, which would make the QoN not overly sensitive (i.e., close to 0 or $\pi$ all the time) for performance optimization.
Fig. \ref{fig:vary_threshold} shows that the appropriate range of $\varepsilon$ for making QoN effective can be $[0.11,0.13]$ (in radian) in our case (experimental setup is in Sec. \ref{sec:experimental_setup}).

\begin{figure}[t]
    \centering
    \setlength{\abovecaptionskip}{0cm}
    \includegraphics[width=0.85\linewidth]{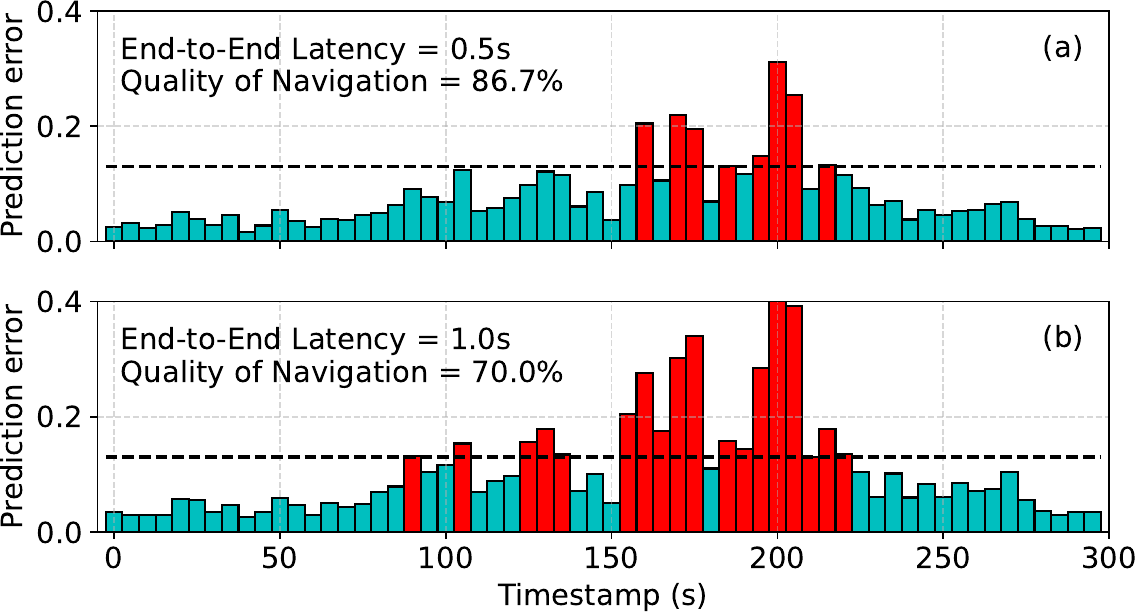}
    \caption{The prediction errors distribution and the corresponding quality of navigation in 300 time-slots when the end-to-end latency is fixed at (a) 0.5s and (b) 1.0s. The dashed line indicates a prediction error threshold of 0.13.}
    \label{fig:error_timestamp}
    \vspace{-0.5cm}
\end{figure}

\begin{figure*}[t]
\centering
\setlength{\abovecaptionskip}{-0.1cm}
\includegraphics[width=0.85\textwidth]{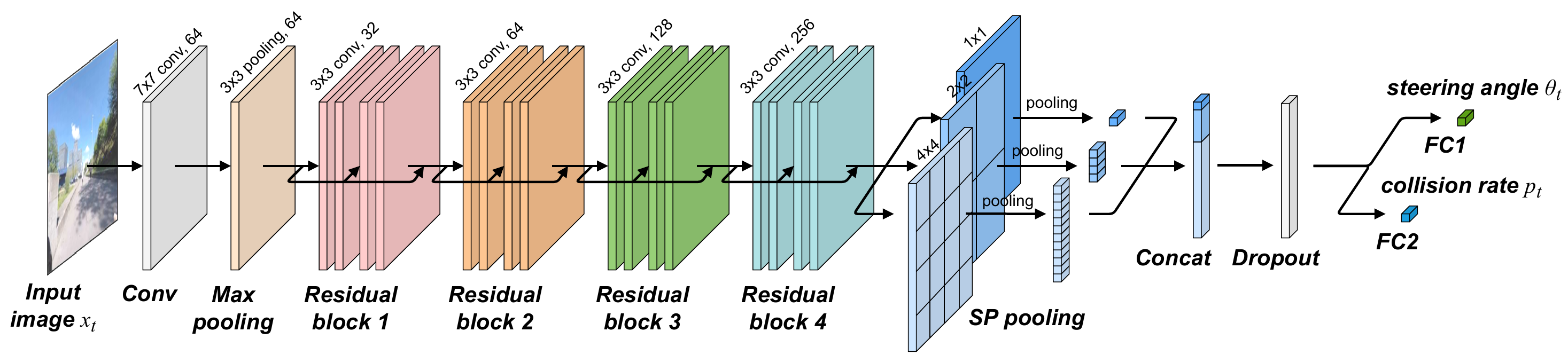}
\caption{We use DroNet as the navigation model of A3D, which inputs a captured image $x_t$ and outputs steering angle $\theta_t$ and collision rate $p_t$ for flight control.
We insert a spatial pyramid (SP) pooling layer to the original DroNet, which enables accepting images of dynamic resolutions.}
\label{fig:navigation_model}
\vspace{-0.5cm}
\end{figure*}

For autonomous drones, QoN can effectively shape navigation performance in terms of latency and accuracy as it inspects the statistics of navigation precision over a given time horizon.
To corroborate that, Fig. \ref{fig:error_timestamp} shows two instances of different decision latency on the same route with the error threshold $\varepsilon = 0.13$ and time window size $\tau = 300$.
In the top subfigure where the latency is fixed at 0.5s, only eight decisions in the period $[150,225]$ break the SLO, while in the bottom subfigure with 1.0s latency, there are 18 failed service events.
Although these two cases share the same navigation model (with the same inference accuracy), their QoNs respectively log at $80.7\%$ and $70.0\%$, demonstrating that our choice of QoN defined in Eq. \eqref{eq:QoN} effectively captures the prediction accuracy and the effects of navigation latency.

We should emphasize that optimizing QoN does not imply minimizing the end-to-end latency directly since we also need to account for the inference quality. For instance, if we always run the lowest input resolution to minimize the latency, it can harm the inference accuracy and produce a large prediction error from the ground truth, leading to a poor QoN.

\begin{figure}[t]
\centering
\setlength{\abovecaptionskip}{-0.1cm}
\includegraphics[width=0.95\linewidth]{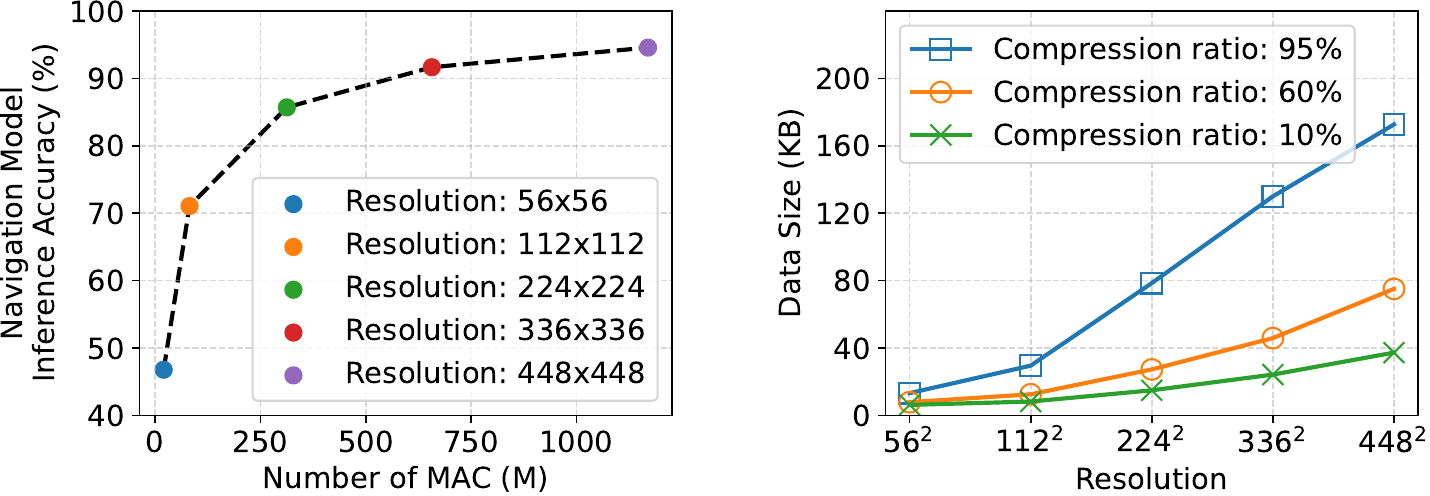}
\caption{The navigation model inference accuracy and the total multiply-accumulate (MAC) operations (left), and the data sizes (right) of input images in different resolutions.}
\label{fig:resolution_relation}
\vspace{-0.4cm}
\end{figure}

\subsection{Design Space of Drone Adaptability}

Viewing navigation as a service allows us to trade inference accuracy for lower latency under the bound of error threshold, and thus improves overall QoN.
To achieve such a goal requires the flexible adaptability of navigation scheduling, where we consider jointly optimizing three key configurations, including input resolution, inference execution location, and image compression ratio.

\textbf{Input resolution.}
Resizing the input image to a lower resolution is a common practice to reduce the computation workload of deep learning models.
Existing systems (e.g., \cite{jiang2018chameleon,wang2020joint}) usually achieve dynamic input resolution by loading a group of models that accept different input sizes and switching the execution target at runtime, which may take a large volume of memory and bring model switching overhead.
To enable dynamic resolution of input images in a lightweight manner, we intend to enhance prevailing models by leveraging the Spatial Pyramid (SP) pooling\footnote{
The SP pooling is originally used only to expand the receptive field, but it enables the model to input arbitrary resolution, and the computational complexity of the model is proportional to the input resolution. Hence, we use SP pooling to achieve the dynamic input resolution without switching models.
} mechanism \cite{he2015spatial}.
Fig. \ref{fig:navigation_model} exemplifies how it is incorporated into DroNet \cite{loquercio2018dronet}: we insert the SP pooling layer in a position where all convolutions are completed.
In our experiments, when using the highest resolution of $448\times448$, A3D's navigation model records merely a tiny accuracy loss of 1\% compared to the original DroNet.
Although SP pooing introduced the execution overhead of three pooling layers, it is negligible in the whole model.

\textbf{Inference execution location.}
Offloading workload to nearby edge servers is another mainstream means to reduce computing latency \cite{zeng2020coedge, ouyang2019cost, zeng2019boomerang}, by utilizing external resources.
In A3D, we regard it as a binary option and will dynamically optimize the selection of inference execution location of the navigation model, i.e. on the drone board or the server.
For simplicity, we assume that there is always an edge server available (e.g., edge servers provided by cellular operators at base stations) for navigation serving during the flight, although the network quality between the drone and edge server may fluctuate.
For the case with multiple servers, we notice that existing literature (e.g, \cite{ouyang2018follow,ndikumana2017collaborative}) has extensively studied strategies for service selection and migration, which can be easily integrated into A3D as supporting modules.

\textbf{Compression ratio.}
To shrink the transmission overhead for offloading, images are usually encoded using lossy compression tools before transfer and decoded as it arrives (JPEG in our implementation).
A3D also makes the compression ratio of this encoding procedure a decision variable to adjust the input image's quality and data size, and therefore tune the tradeoff between inference accuracy and end-to-end latency.

\begin{figure}[t]
\centering
\setlength{\abovecaptionskip}{-0.1cm}
\includegraphics[width=0.95\linewidth]{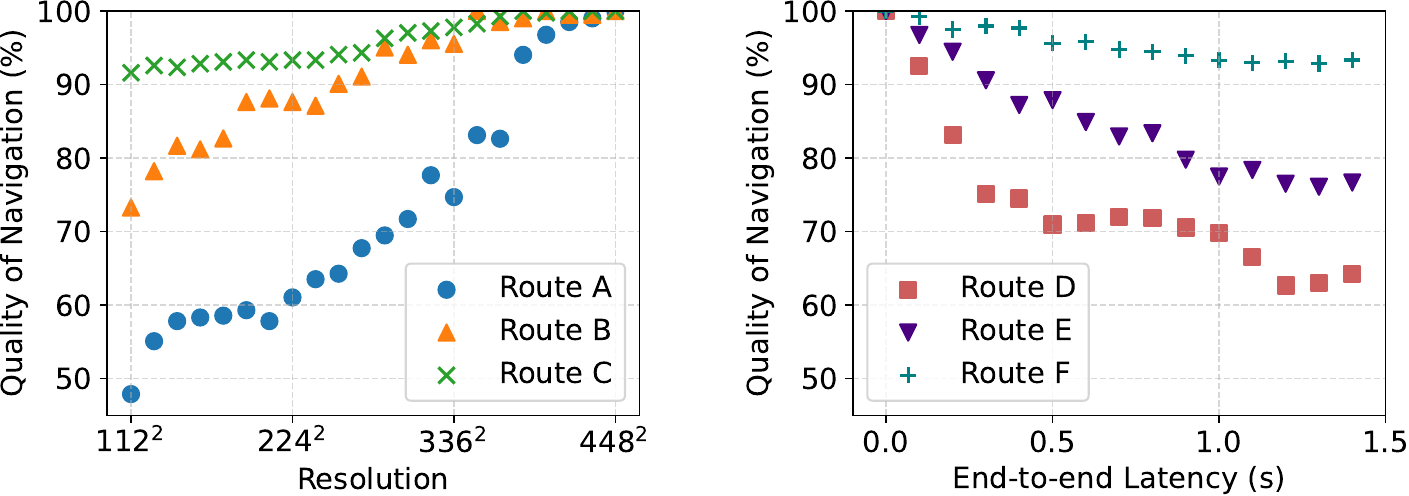}
\caption{The measured quality of navigation varies in different routes with respect to the changes of resolution (left) and end-to-end latency (right).}
\label{fig:dynamic}
\vspace{-0.4cm}
\end{figure}

\subsection{Challenges of Scheduling Adaptive Navigation}
\label{label:challenge}

Given the above design space and serviceable objective, achieving adaptive navigation in high performance is non-trivial, following three critical challenges.

\textbf{(1) Composite optimization objective.}
QoN is a composite target blending both inference accuracy and latency, while optimizing these two metrics separately is usually in conflict under resource constraints. 
Reducing latency is often at the expense of accuracy, and improving accuracy often requires enduring higher latency. 
To strike a good tradeoff requires a careful analysis of their relationship, which is challenging.

\textbf{(2) Complex nexus of schedulable configurations.}
The impact of three schedulable dimensions does not independently act on the targeted QoN objective, but exhibit in an assorted manner.
For example, centering on the input images, Fig.\ref{fig:resolution_relation} shows the effect of input resolution and compression ratio dimensions: 
the decrease in input resolutions can well reduce the computing workload in total multiply-accumulate (MAC) operations (left subfigure) and the data sizes (right subfigure), both of which encourage lower latency, and selecting a smaller compression ratio can further magnify that. 
However, they come at the price of accuracy drops, and if the resolution is too small (e.g. $56\times56$), the accuracy can be unusable and QoN suffers.

\textbf{(3) Dynamic environmental information.}
The challenge of adaptability also lies in the dynamic edge environment with respect to 1) networking conditions, 2) routes' navigation difficulty, and 3) environmental scenes' changes.
Particularly, we illustrate the latter two factors using measurements on different routes.
In Fig. \ref{fig:dynamic}(left), we observe that QoN's sensitivity to different resolutions varies in different routes, indicating that the inference precisions of their corresponding input image also vary.
In Fig. \ref{fig:dynamic}(right), the pattern is analogous where the achieved QoN data points are in different levels under the same latency premise in different routes.
Overall, as the drone keeps flying, the physical surroundings are changing, requiring conscious environmental awareness for adaptive scheduling.

\begin{figure}[t]
\centering
\setlength{\abovecaptionskip}{-0.1cm}
\includegraphics[width=0.95\linewidth]{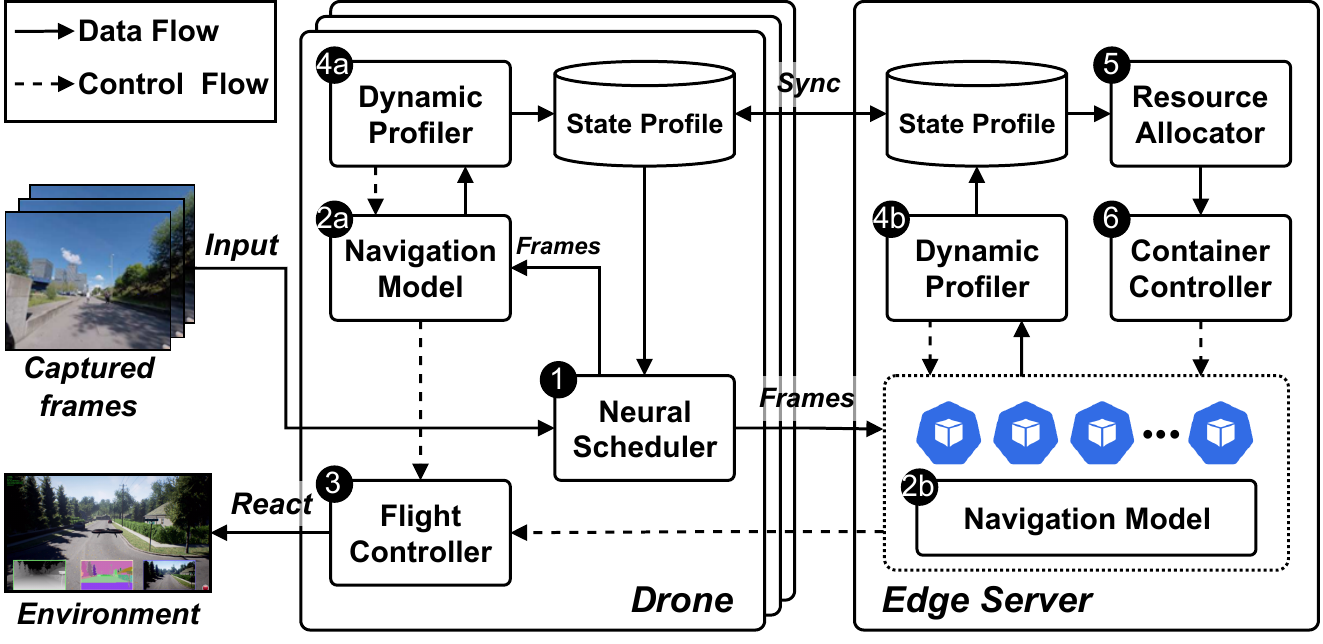}
\caption{A3D architecture overview. Given a series of captured images, the neural scheduler decides an execution location, and accordingly adjusts the input image and transfers the frames to the navigation model for flight control.}
\label{fig:architecture}
\vspace{-0.4cm}
\end{figure}

\section{System Overview}
\label{sec:overview}

To address the above challenges, we propose A3D, an adaptive scheduling framework across drones and edge servers for high-quality autonomous navigation tasks. 
Fig. \ref{fig:architecture} shows the architecture of A3D. 
First, the onboard computing device acquires the images captured by the camera and passes them to the neural scheduler (\ding{202}).
The scheduler is responsible for scheduling a system configuration in a design space comprised of image resolution, inference execution location, and the compression ratio, targeting maximizing the QoN performance.
In particular, the input image is resized and compressed (if needed) according to the determined image resolution and compression ratio, fed as the input to the navigation model (on the board or the edge server).
If the execution location is instantiated as the edge server according to the configuration, the compression ratio of the input image is subsequently adjusted to encode the images, and thereafter sent to the server for inference (\ding{203}, Sec. \ref{sec:neural_scheduler} ). 
The navigation model on the server runs in a containerized environment, and is managed by the container controller (\ding{207}). 
It outputs navigation decisions and sends them to the flight controller (\ding{204}), which in turn forwards the flight commands to the drone following the control loop in Fig. \ref{fig:control_loop}. 
During the runtime, the dynamic profiler (\ding{205}) continuously monitors system profiles including bandwidth $b$, server computing resources $s$ and navigation model output $\theta , p$. 
To support concurrent multi-drone serving, a resource allocator (\ding{206}, Sec. \ref{sec:support_multiple_drones}) is further developed to intelligently assign proper computing resources to containers (corresponding to individual drones).
The dynamic profiler and the state profile are deployed on both the onboard device and the server, since the navigation model may be executed alternately on either side.
As their profilers only have access to a portion of the environmental information, the two state profiles are synchronized periodically to ensure data integrity.

\begin{figure}[t]
\centering
\setlength{\abovecaptionskip}{-0.1cm}
\includegraphics[width=0.95\linewidth]{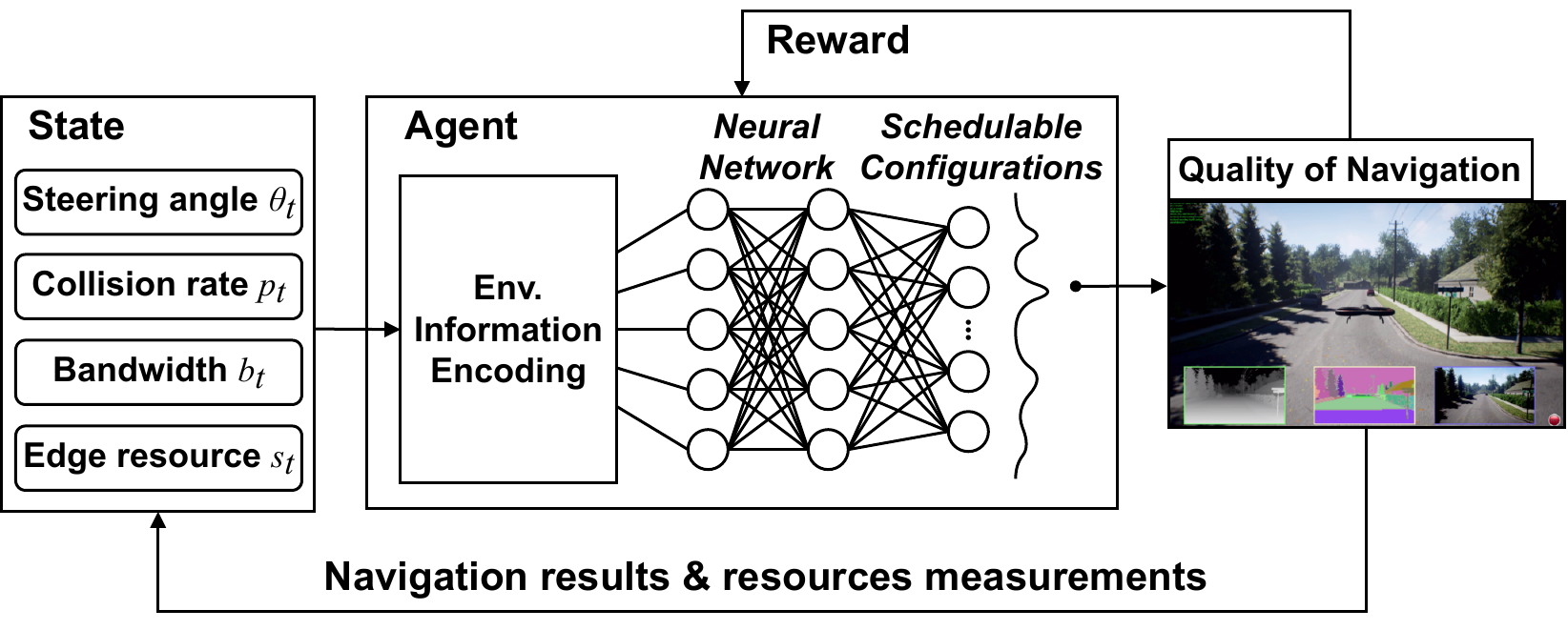}
\caption{In A3D's DRL-based neural scheduler, an agent observes the navigation states to decide a scheduling action on the flight environment and receives a reward based on the quality of navigation. The agent uses environmental information encoding to model the environment complexity and dynamics.}
\label{fig:drl}
\vspace{-0.4cm}
\end{figure}

\section{Neural Adaptive Scheduler}
\label{sec:neural_scheduler}

Scheduling navigation for real-time, adaptive, and efficient performance is intractable, provided challenges discussed in Sec. \ref{label:challenge}.
What's worse, the irregularity and non-smoothness of the targeted QoN objective make the problem non-convex and hard to be analytically expressed, leaving existing mathematical methodology unavailable for efficient optimization.
Therefore, instead of characterizing connections between variables and QoN individually, A3D treats the entire system as a black box and learns to solve the optimization using a DRL-based neural scheduler.
Beyond merely applying off-the-shelf DRL algorithms, we design an environmental information encoding mechanism to reshape the state features, which turn out to be a better state abstraction for accelerating the training convergence and promoting the obtained policy.

\subsection{Framework Overview}
A3D's RL framework (Fig. \ref{fig:drl}) is general and can be applied to a variety of navigation objectives.
Specifically, it intends to schedule configurations, observe the outcome, and provide the agent (neural network) with a reward after each action.
We refer to each component as state, action, and reward, defined in detail as follows.

The \textbf{state} consists of the observable environmental information at time $t$, including the output of the navigation model (steering angle $\theta_t$ and collision rate $p_t$), bandwidth $b_t$ between the drone and the edge, and edge computing resource $s_t$ allocated by the server (measured in available CPU cores). 
In summary, the state space is defined as $\mathcal{S}=\langle \theta_t,p_t,b_t,s_t \rangle$.

The \textbf{action} should be consistent with the schedulable configurations, i.e. input resolution $r$, inference execution location $o$ and image compression ratio $j$.
Namely, the action space is $\mathcal{A}=\langle r,o,j \rangle$. 
To reduce the training difficulty and accelerate the convergence, we discretize the action space, where $r\in\{448\times448,224\times224,112\times112\}$, $o\in\{0,1\}$ ($0$ for drone board and $1$ for edge server), and $j\in\{95,60,10\}$.
Note that $j$ and $o$ are coalescent as image compression is available if and only if offloading is chosen ($o=1$).
All actions are encoded in a zero-one vector.

The \textbf{reward} is exactly the optimization objective QoN $\mathcal{Q}$.
In the training process, we measure $\mathcal{Q}$ by Eq. (\ref{eq:QoN}) after every DRL step. The size of the time window $\tau$ is equal to the length of a DRL step, which is set to 5s in our case, such that the QoN is averaging approximately $17$ times of navigation inferences for each policy update in the DRL training.
Note that the computation of QoN at the runtime is not required, since the DRL agent will output the action based on the state directly.

\subsection{Environmental Information Encoding}
\label{label:encoding}
Applying neural networks as the agent enables the DRL scheduler to possess the ability of fitting nonlinear functions, and thus can learn the relationships between the variables and the objective, addressing Challenges (1) and (2).
However, to support the scheduler to process environmental information dynamically, i.e. Challenge (3), it requires further enhancement in identifying input difficulties, and we develop an Environmental Information Encoding (EIE) module to deal with that.

The core of EIE is two knobs that reflect the properties of captured images.
The first is \textit{environment complexity} $c$ that characterizes how sensitive the QoN is to the change of input resolutions.
Formally, we define $c$ as a weighted sum of the navigation decisions' variants in different resolutions:
\begin{align}
    c = |\theta_{\text{high}}-\theta_{\text{low}}| + \alpha |p_{\text{high}}-p_{\text{low}}|, \label{eq:env_complexity}
\end{align}
where $\alpha$ is a hyper-parameter that keeps $|\theta_{\text{high}}-\theta_{\text{low}}|$ and $|p_{\text{high}}-p_{\text{low}}|$ at the same order of magnitude, and the subscripts indicate results corresponding to images in the highest and lowest resolutions, respectively.
In A3D, we use a profile-based approach to measure $c$ at system idle time: first record the navigation model's outputs with images in $448\times448$ and $112\times112$, then calculate $c_t$ according to Eq. (\ref{eq:env_complexity}).

The second is \textit{environment dynamics} $d$ that characterizes how rapidly the content of captured images changes.
This indicator induces the expiration time for the current navigation decision, implying the urgency of optimizing inference latency.
We therefore define $d$ using the distributional divergence of $\theta$ and $p$ within the latest navigation epoch:
\begin{align}
    d = \sigma(\theta) + \beta\sigma(p), \label{eq:env_dynamic}
\end{align}
where $\sigma(\cdot)$ reckons the standard deviation and $\beta$ is a hyper-parameter.
The rationale behind Eq. (\ref{eq:env_dynamic}) is to regard the model output as a descriptor of the image, where the degree of model output's variation can induce the degree of image content's variation, i.e. environmental changes. 
Estimating $d$ only needs to record navigation decisions at runtime and does not introduce additional overhead.

\begin{figure}[t]
\centering
\setlength{\abovecaptionskip}{-0.1cm}
\includegraphics[height=3cm]{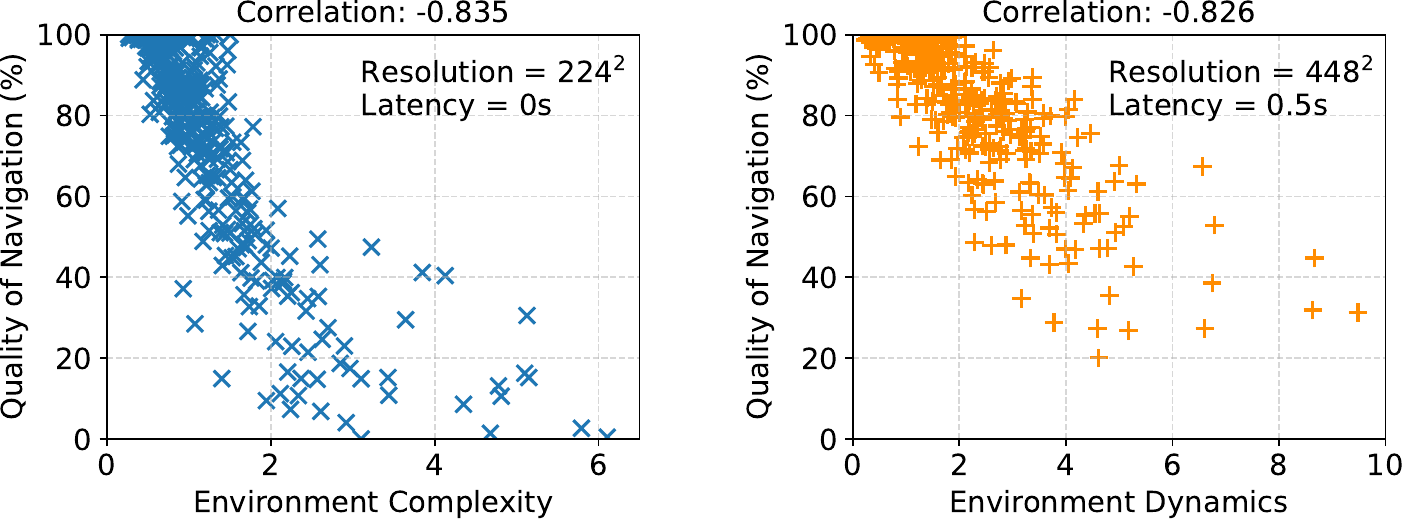}
\caption{The quality of navigation declines as the environment complexity (left) and the environment dynamics (right) increase. Their Pearson correlation coefficients are -0.835 and -0.826, respectively. }
\label{fig:env_complexity_dynamic}
\vspace{-0.6cm}
\end{figure}

We verify the effectiveness of the above two definitions on the Mid-Air dataset \cite{fonder2019mid}, by recording the environment complexity and dynamics, as well as the achieved QoN, in every-5s time slots.
Fig. \ref{fig:env_complexity_dynamic} shows the data points, where we fix the resolution at $224\times224$/$448\times448$ and latency at 0s/0.5s, respectively. 
Visualized results show the evident correlation between environment complexity $c$ and the degradation of QoN: the larger $c$ is, the more complex the environment is, and thus the smaller value QoN logs.
The same pattern also holds for environment dynamics $d$, demonstrating their ability in shaping environment properties. 
Statistically, the Pearson correlation coefficients are -0.835 and -0.826, respectively, indicating a strong negative tendency between the targeted QoN and $c$ ($d$).
Hence, with the EIE mechanism, state $\mathcal{S}$ of the DRL agent at time $t$ is refined as $\langle c_t,d_t,b_t,s_t \rangle$ without directly using $\theta$ and $p$.

\subsection{Training}
A3D's neural scheduler employs the Actor-Critic algorithm (A2C) \cite{konda2000actor} for training, which combines a value-based algorithm and a policy gradient-based algorithm.
We select it because of its advantages of low inference latency and fast training convergence as we will show later in Sec. \ref{sec:eval_scheduler}.

To speed up the training process, we construct a numerical simulation environment to train the DRL agent.
We use Mid-Air \cite{fonder2019mid}, a drone flight video streaming dataset that lasts for 80 minutes and contains about 420,000 frames covering multifarious weather conditions and environments.
We employ the publicly-available wireless bandwidth traces dataset HSDPA\cite{riiser2013commute} to simulate the fluctuations of networking conditions during flight.

We use the Jetson Nano as an onboard computing device to measure the computing latency of the navigation model for different resolution inputs. We assume that these latency data are constant at runtime, and use the measurements as runtime data to construct a drone simulation environment. 

Furthermore, we use offline data to speed up training, generating predictions of the navigation model $\theta,p$ for all 420,000 frames using all scheduling decisions defined in the action space beforehand and recording in a table. 
During the DRL training, we directly look up corresponding results from the table and consequently save the navigation inference time.
By doing so, our simulation allows the DRL agent to ``experience'' 80 minutes of flight in 10 minutes.
\vspace{-0.2cm}

\section{Supporting Multiple Drones}
\label{sec:support_multiple_drones}

The neural scheduler introduced in Sec. \ref{sec:neural_scheduler} allows individual drones to adaptively decide whether to resort to the edge server's assistance for accurate navigation.
However, while a swarm of drones flies around and separately sends offloading queries, the edge server is obliged to serve multiple DNN models and infer their navigation decisions.
In this circumstance, existing literature usually considers a buffering strategy, which accepts serving queries in a queue and processes them with exclusive, sufficient resources in a streaming manner.
Although it can substantially alleviate resource contention, the delay and overhead caused by buffering are problematic.
On the one hand, the buffering process necessarily prolongs the end-to-end latency perceived by the drone (when the offloading decision is applied), which severely damages the responsiveness and efficacy of the edge-assisted solution.
On the other hand, learning a DRL model (neural scheduler) to assure a steady, content reward toward the QoN objective becomes much more challenging given the buffering delay, which is hard to be predicted and maintained.
Therefore, we instead leverage a concurrent serving principle at the edge server that adaptively assigns proper edge resources for individual drones and serves them simultaneously.
In what follows, we will explain the proposed network-aware resource allocation algorithm in detail.
\vspace{-0.2cm}

\subsection{Resource Allocation for Multiple Drones} 

The functionality of edge resource allocation is accomplished by the resource allocator (Fig. \ref{fig:architecture} \ding{206}) at the edge server, operating upon the container controller.
Its objective is to maximize the global drone performance, quantified by the average QoN of all served drones.
To schedule a proper resource allocation solution is non-trivial, given the following challenges.
First, the resource demand for navigation may differ across individual drones, since their system configurations on image resolution, execution location, and compression ratio may vary on the fly.
This attributes to many factors, e.g., their captured images are different when flying at different routes and heights, and their local computing resources and networking conditions are also diverse.
Second, the actual demand for edge resources is unknown apriori, and is implicitly intertwined with the allocated volume of edge resources. 
To be more specific, the computational workload at the edge server highly depends on the system configuration $\mathcal{A}$ (e.g., the image resolution) determined by the neural scheduler, which contrariwise relies on the input state $\mathcal{S}$ that comprises the allocated edge resource $s$.
Third, serving a group of drones concurrently may lead to critical resource contention for navigation model inference, given that edge servers are typically with a relatively moderate scale of computing resources (compared to the powerful cloud datacenters).
Such resource shortage can lead to serious performance degradation, which may conversely hinder the drones' QoN.

\begin{figure}[t]
\centering
\setlength{\abovecaptionskip}{-0.1cm}
\includegraphics[height=3cm]{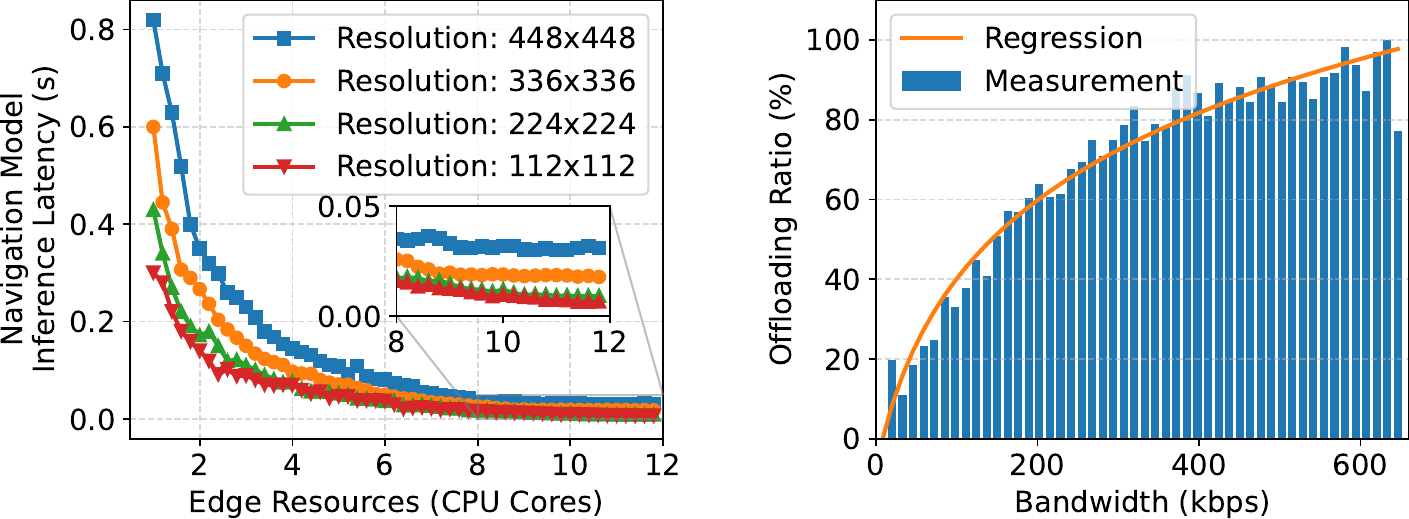}
\caption{Left: The navigation model inference latency as a function of available edge resources. Right: The offloading ratio approximately grows in a logarithmic trajectory as the bandwidth increases.}
\label{fig:multi_drone_motivation}
\vspace{-0.6cm}
\end{figure}

To explore how allocated resources impact flying performance, we examine the navigation model inference latency on the edge server by varying the assigned CPU cores to the corresponding container (in a granularity of 0.1 virtual CPU cores).
Fig. \ref{fig:multi_drone_motivation}(left) depicts the results with input images in different resolutions, where we remark three observations.
First, with more resources the inference latency gradually lowers, showing the clear benefit of resource replenishment for all resolution settings.
When increasing CPU cores from 1 to 10, the navigation model inference achieves at most 26.5$\times$ speedup (for $448\times448$ resolution).
Second, inference queries with different input image resolutions exhibit differentiated sensitivity to the resource variation.
A higher-resolution workload (e.g., 448$\times$448) gains a larger latency reduction with the same resource supplement.
Third, the performance gap between the resolution settings shrinks as the edge resources become more abundant.
If the CPU cores are adequately ample (e.g., $>$10), the inference latency even appears a convergence and the benefit of adding more cores marginally diminishes.
This inspires us that allocating resources in the middle region (e.g., [4,8] in Fig. \ref{fig:multi_drone_motivation}(left)) can maximize the edge resources utilization.
Besides, we can assert that a trivially random or equal allocation cannot sufficiently meet the service requirement, where an on-demand solution that aligns the need for drone queries and edge computing resources is desired.

Designing such an on-demand solution, however, necessitates an effective estimation of drones' reliance on the edge server, which is hard to predict accurately.
Instead of applying a precise but prohibitively expensive estimation approach, we observe that the networking condition, i.e., bandwidth, can be utilized as a general indicator to reflect drones' reliance on edge.
The rationale behind is that with higher communication bandwidth, the drone is more likely to offload its computation to the edge server.
To validate that, we experiment with the proposed neural scheduler by adjusting the bandwidth $b$ and logging the average offloading ratio within a period, which yields the results in Fig. \ref{fig:multi_drone_motivation}(right).
We witness that the higher the bandwidth, the higher possibility the drone would offload its workload.
More surprisingly, the recorded data points of the offloading ratios exhibit a logarithmic tendency (plotted in the curve in Fig. \ref{fig:multi_drone_motivation}(right)), indicating a logarithmic regression model can approximately map the profile-friendly networking conditions to the allocation-related offloading ratio.

Summarizing the above observations motivates us to design an on-demand strategy that leverages the bandwidth as a knob and allocates edge resources to match the drones' demand.

\begin{algorithm}[t] 
\caption{Network-aware resolution allocation algorithm} 
\label{algo:resource_allocation} 
\begin{algorithmic}[1] 
    \REQUIRE ~~\\ 
    $\langle b_1, b_2, \cdots, b_n \rangle$: The measured bandwidths between drones and the edge servers \\
    $\mathcal{R}$: Trained regression model that maps bandwidth to an offloading ratio \\
    $h, l$: The upper and lower bounds for resource allocation \\
    \ENSURE ~~\\ 
    $\langle s_1, s_2, \cdots, s_n \rangle$: The allocated edge resources for drones
    \STATE \textit{/* - - - Initialization - - - */}
    \STATE $\langle f_1, f_2, \cdots, f_n \rangle \gets \mathcal{R} (\langle b_1, b_2, \cdots, b_n \rangle)$
    \STATE Calculate $s_i$ according to Eq. (\ref{eq:init_alloc})
    \STATE \textit{/* - - - Bounded reallocation - - - */}
    \STATE Construct a set $\Psi$ with the elements $s_i$ in $\langle s_1, s_2, \cdots, s_n \rangle$ such that $s_i > h$ and assign $s_i \gets h$
    \STATE Calculate the resource surplus $S^+$ by Eq. (\ref{eq:resource_surplus})
    \STATE Construct a set $\Phi$ with the elements $s_i$ in $\langle s_1, s_2, \cdots, s_n \rangle$ such that $s_i < l$ and assign $s_i \gets l$
    \STATE Calculate the resource shortage $S^-$ by Eq. (\ref{eq:resource_shortage})
    \STATE $\Theta \gets \langle s_1, s_2, \cdots, s_n \rangle - \Psi - \Phi$
    \WHILE{\text{True}}
        \STATE $\Delta S \gets S^+ - S^-$
        \IF{$\Delta S < 0$}
            \STATE Find the least element $s_{\text{min}}$ in $\Phi$ and set $s_{\text{min}} \gets 0$
            \STATE $S^- \gets S^- - l$
        \ELSE
            \STATE Assign $\Delta S$ to the elements in $\Theta$ proportionally
            \STATE Break
        \ENDIF
    \ENDWHILE
    \RETURN $\langle s_1, s_2, \cdots, s_n \rangle$
\end{algorithmic}
\end{algorithm}

\subsection{Network-Aware Resource Allocation Algorithm} 

The key idea of the proposed resource allocation algorithm is a two-phase scheduling: first initialize a resource allocation solution via the estimated offloading ratio, and next refine it by aligning in a proper interval.
Algorithm \ref{algo:resource_allocation} shows the procedure, where its input includes 1) the measured bandwidths $\langle b_1, b_2, \cdots, b_n \rangle$ between drones and the edge server, 2) the trained regression model $\mathcal{R}$ that can map a given bandwidth $b_i$ to the estimated offloading ratio $f_i$, and 3) the operator-defined upper bound $h$ and lower bound $l$ for resource reallocation.
The expected output is the allocated edge resources $\langle s_1, s_2, \cdots, s_n \rangle$ for individual drones.

Algorithm \ref{algo:resource_allocation} begins at the first phase that calls the regression model $\mathcal{R}$ to estimate the offloading ratio $\langle f_1, f_2, \cdots, f_n \rangle$ for all drones, taking the profiled bandwidth as input.
With these estimations, we initialize a preliminary allocation in proportion to the drones' offloading possibilities, using Eq. (\ref{eq:init_alloc}):
\begin{align}
    s_i = \lambda \frac{f_i}{\sum_{j=1}^{n} f_j}, \label{eq:init_alloc}
\end{align}
where $\lambda$ is the amount of available resources at the edge server.
Next, the algorithm enters the second phase for allocation refinement.
In particular, it first finds the elements in the current solution $\langle s_1, s_2, \cdots, s_n \rangle$ that have values out of the interval $[l, h]$.
For the elements with values higher than the upper bound $h$, we collect them in a set $\Psi$ and reassign their values exactly with $h$.
Meanwhile, we calculate the resource surplus $S^+$ derived from the above reassignment by Eq. (\ref{eq:resource_surplus}) (line 5-6).
Similarly, for the elements with values smaller than the lower bound $l$, we repeat the same procedure with Eq. (\ref{eq:resource_shortage}) and obtain a set $\Phi$ and the resource shortage $S^-$ (line 7-8).
We count the unchanged elements by filtering the current solution with $\Psi$ and $\Phi$, denoted in a set $\Theta$.
\begin{align}
    S^+ &= |\textstyle \sum_{s_j \in \Psi}s_j - |\Psi|\cdot h|, \label{eq:resource_surplus} \\
    S^- &= |\textstyle \sum_{s_j \in \Phi}s_j - |\Phi|\cdot l|. \label{eq:resource_shortage}
\end{align}

The algorithm then dives into an iteration that intends to generate a valid allocation after the above reassignment.
To gauge how much resource is remained, we reckon the difference between $S^+$ and $S^-$ and obtain $\Delta S$.
If $\Delta S < 0$, the allocation meets a resource deficit.
To ensure a valid solution, we select the least element in $\Phi$ and reset it to 0, which implies that the edge server will not allocate resources for the corresponding drone.
The rationale behind is that with fewer edge resources the drone is less possible to offload its workload, and even if it decides an offloading configuration, the inference latency on the edge side will be too high to satisfy the navigation service (as in Fig. \ref{fig:multi_drone_motivation}(left)).
After dropping this drone's service, its originally owned resource is released and can be used for further reallocation (in another iteration of the loop).
If $\Delta S \geq 0$, there are still spare resources available for allocation, so we assign $\Delta S$ to the elements in $\Theta$ in proportion to their offloading ratios and break the loop (line 16-17).
The algorithm terminates by returning the final allocation $\langle s_1, s_2, \cdots, s_n \rangle$.

Algorithm \ref{algo:resource_allocation} takes $O(n)$ time complexity with $n$ drones.
Given that the amount of drones in a swarm is typically several or tens, the algorithm is lightweight and can run efficiently, which allows fast and agile edge resources scheduling during the edge server's runtime.
The selection of the bounds $h$ and $l$ is given by the system operator, which can be flexibly tuned to accommodate the navigation model's performance, the edge server's capability, as well as the input image's complexity.

\section{Implementation}
\label{sec:implementation}
With all the above designs, we explain our implementation in this section, in terms of the proof-of-concept prototype and the simulation environment.
\vspace{-0.2cm}

\subsection{Prototype Implementation}

\begin{figure}[t]
\centering
\setlength{\abovecaptionskip}{-0.1cm}
\includegraphics[height = 4cm]{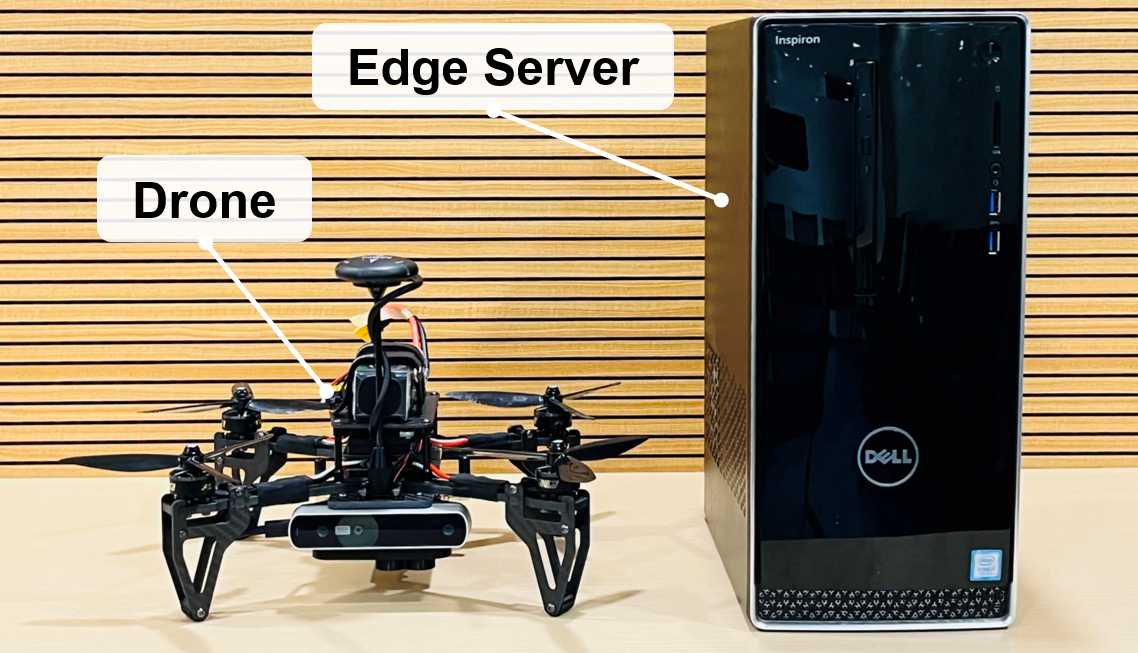}
\caption{The drone and the employed edge server used in our prototype implementation, communicated via a wireless connection. The drone equips with an UP Core as its core processor.}
\label{fig:prototype_hardware}
\vspace{-0.2cm}
\end{figure}

We implement the hardware platform of A3D as shown in Fig \ref{fig:prototype_hardware}: we select the Holybro PX4 Vision Development Kit, a mature commercial product widely used by the community, as the drone. The kit contains a near-ready-to-fly carbon-fiber quadcopter equipped with a Pixhawk 4 flight controller, UP core companion computer, and the Occipital Structure Core depth camera sensor. The workstation equipped with an Intel Xeon(R) W-2145 CPU is not only emulated as the edge server but also functioned as the ground station of the flying drone. The drone kit provisions the external antenna to enable the wireless connection between the drone and the ground station, and the maximum bandwidth of the WiFi connection between the companion computer and the edge server is around 54 Mbps by means of actual measurement. It is noteworthy that we abandon the integrated PX4 obstacle avoidance in this vehicle and we mainly exploit the potential of the captured RGB images rather than the RGBD images.

\begin{figure}[t]
\centering
\setlength{\abovecaptionskip}{-0.1cm}
\includegraphics[width=0.9\linewidth]{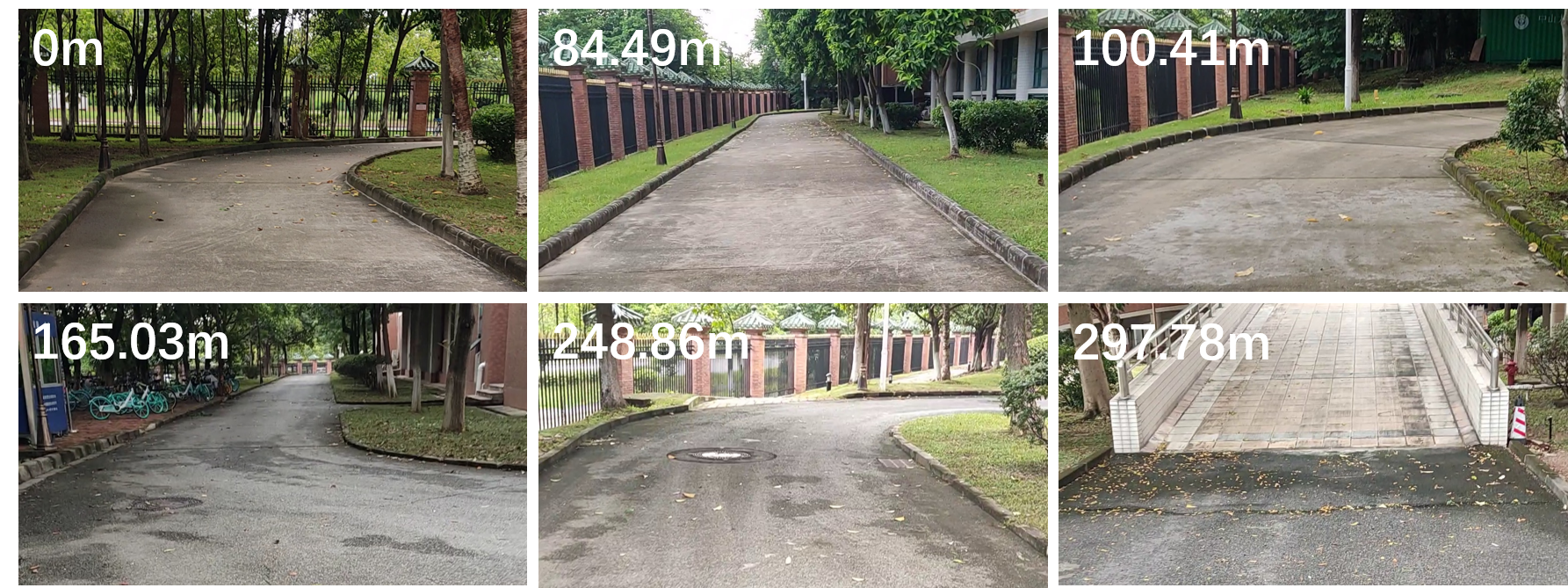}
\caption{The 300m real-world route used in our prototype experiment locates at the campus.}
\label{fig:prototype_route}
\vspace{-0.5cm}
\end{figure}

We utilize the drone to conduct the real-scenario autonomous navigation on a campus route illustrated in Fig \ref{fig:prototype_route}. This route is composed of several straights and turns, and the main pavement is obvious and flanked by green belts aside. The total distance of the path is approximately 300m and some important turns and spots are shown in Fig \ref{fig:prototype_route}. The PX4 flight controller provides the offboard flight mode to assign the control of the vehicle to the companion computer\cite{px4guide}. The companion computer can transform the expected flight instructions into the MAVLink message to control the drone at the hardware level.

\begin{figure}[t]
\centering
\setlength{\abovecaptionskip}{-0.1cm}
\includegraphics[height = 4cm]{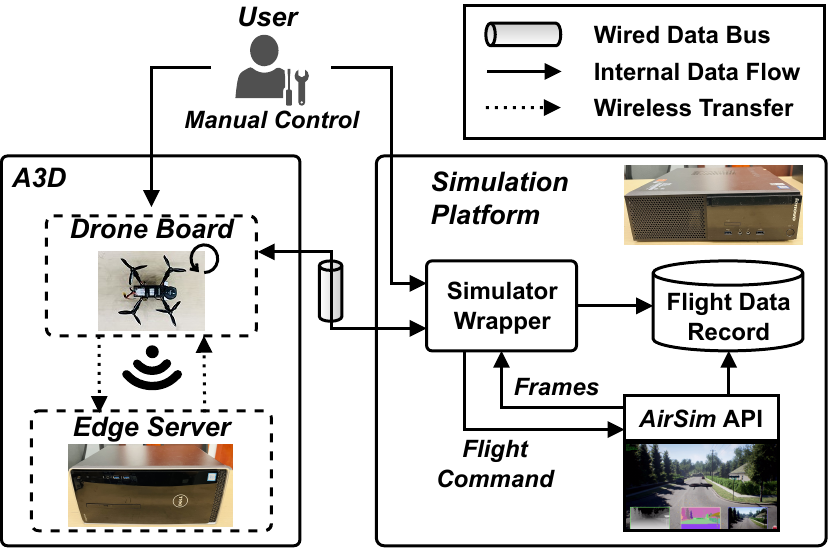}
\caption{A3D integration with AirSim simulator. The prototype connects the drone board (Jetson Nano) and the simulation platform with a data bus and develops a wrapper to manage all simulation data through AirSim API.}
\label{fig:simulator}
\vspace{-0.2cm}
\end{figure}

\begin{figure}[t]
\centering
\setlength{\abovecaptionskip}{-0.1cm}
\includegraphics[width=0.9\linewidth]{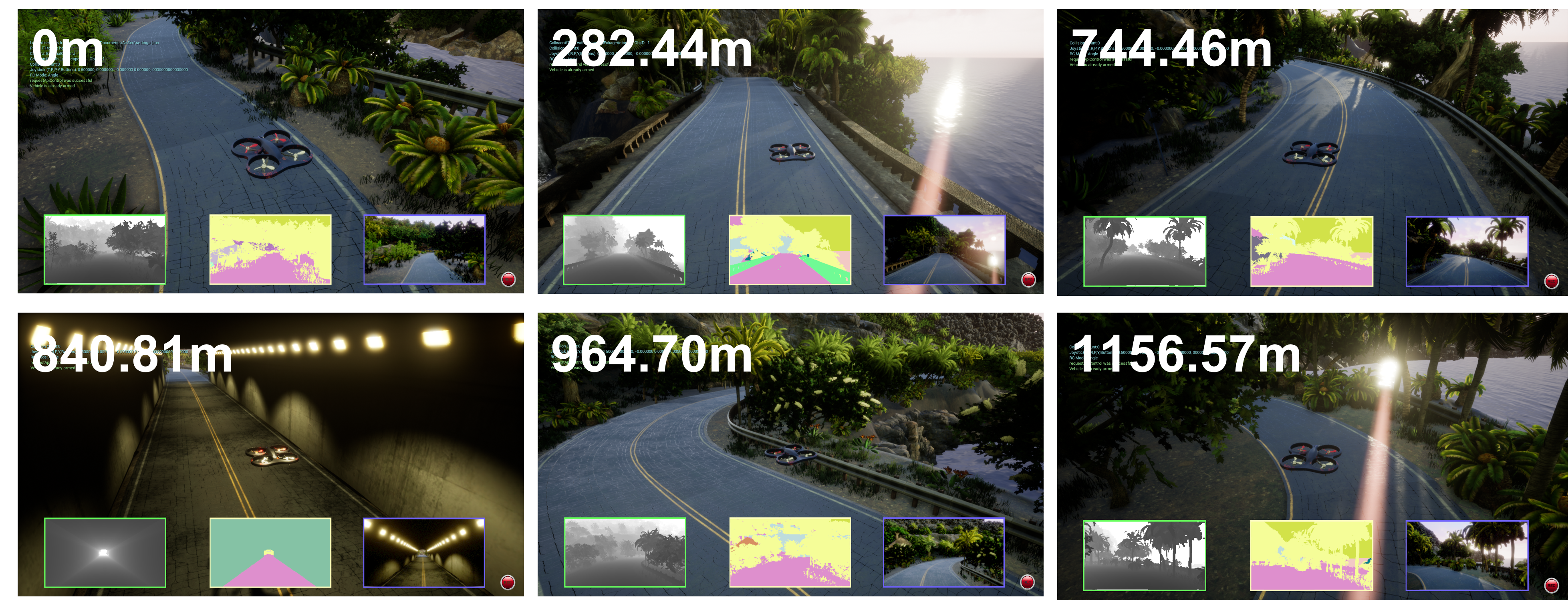}
\caption{The used 1200m coastline route in our simulation covers various types of scenes including straight roads, curves, and tunnels.}
\label{fig:coastline_scene}
\vspace{-0.5cm}
\end{figure}

\begin{figure*}[t]
    \centering
    \setlength{\abovecaptionskip}{0cm}
    \subfigure[Flight distance as a function of end-to-end latency in campus route.]{
        \begin{minipage}[t]{0.22\textwidth}
            \centering
            \setlength{\abovecaptionskip}{-0.1cm}
            \includegraphics[height=3.1cm]{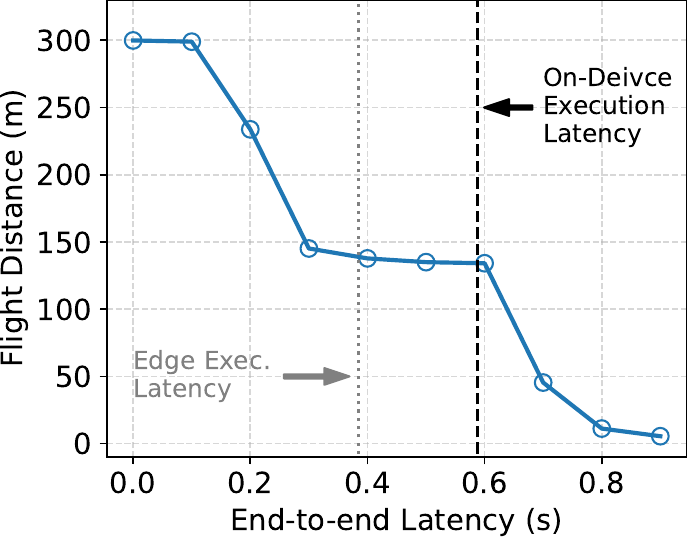}
            \label{fig:prototype_distance_latency}
            \vspace{-0.5cm}
        \end{minipage}
    }
    \
    \subfigure[CDF of navigation model prediction error within the flight period.]{
        \begin{minipage}[t]{0.22\textwidth}
            \centering
            \setlength{\abovecaptionskip}{-0.1cm}
            \includegraphics[height=3.1cm]{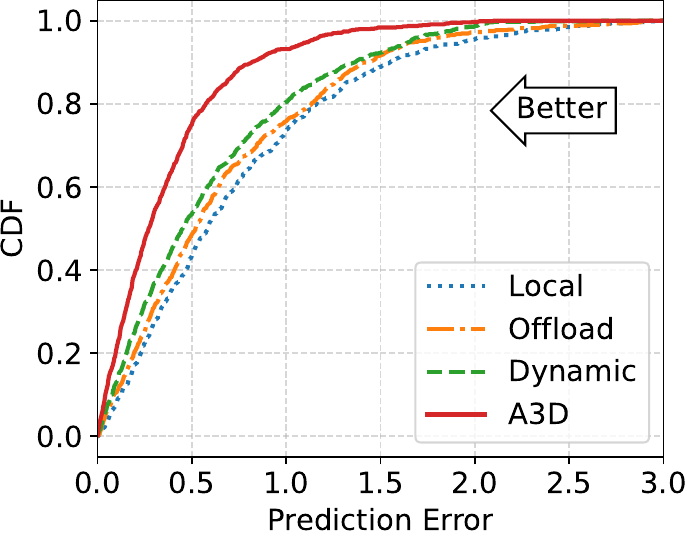}
            \label{fig:prototype_accuracy_distribution}
            \vspace{-0.5cm}
        \end{minipage}
    }
    \
    \subfigure[The distribution of end-to-end latency within the flight period.]{
        \begin{minipage}[t]{0.22\textwidth}
            \centering
            \setlength{\abovecaptionskip}{-0.1cm}
            \includegraphics[height=3.1cm]{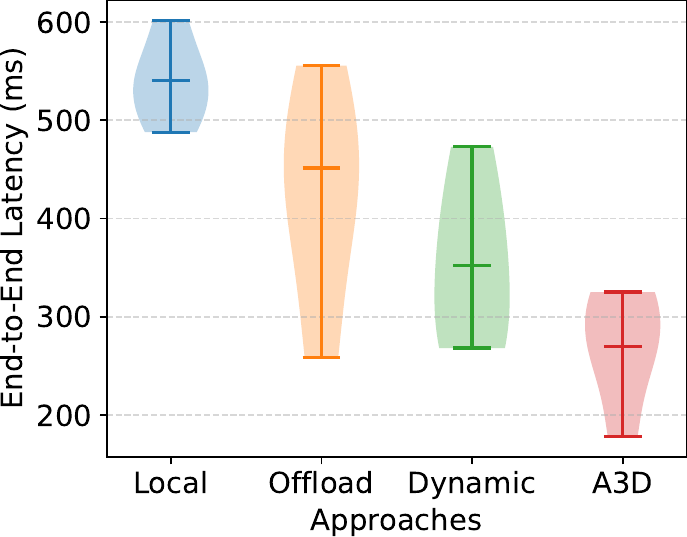}
            \label{fig:prototype_latency_distribution}
            \vspace{-0.5cm}
        \end{minipage}
    }
    \
    \subfigure[QoN of the prototype with varying maximum drone speed.]{
        \begin{minipage}[t]{0.22\textwidth}
            \centering
            \setlength{\abovecaptionskip}{-0.1cm}
            \includegraphics[height=3.1cm]{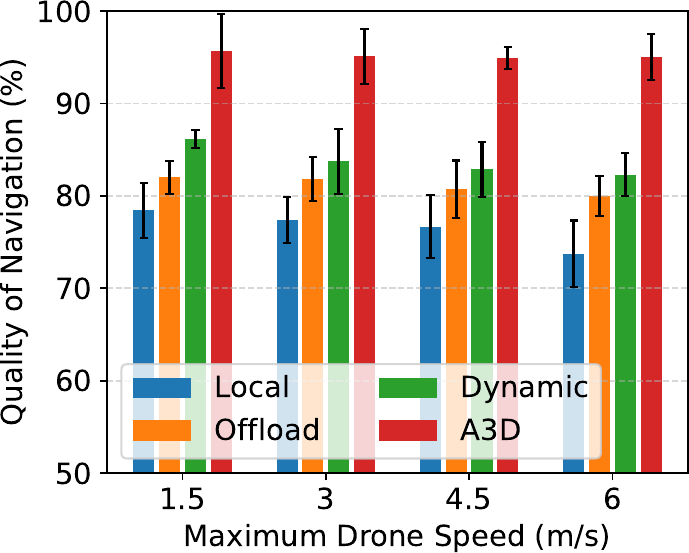}
            \label{fig:prototype_qon}
            \vspace{-0.5cm}
        \end{minipage}
    }
    \caption{Prototype evaluation results.  }
    \label{fig:prototype_experiments}
    \vspace{-0.5cm}
\end{figure*}

\vspace{-0.2cm}

\subsection{Simulation Implementation}
To make a thorough evaluation with more settings, we use the AirSim\cite{shah2018AirSim} platform for simulation. The benefits of simulation lie in that it has no damage to the equipment and high reproducibility of experiments. AirSim is developed by Microsoft based on Unreal Engine 4 (UE4). 
AirSim provides APIs to interact with drones in the simulator. Specifically, the \textit{simGetImages} method is used to obtain the camera images, the \textit{simGetVehiclePose} method is used to obtain the drone's pose, and the \textit{moveByVelocityZAsunc} method is used to specify the drone's flight speed and turn angle. In addition, AirSim provides functions to change the weather conditions and sun angle to simulate various environmental conditions.

Fig. \ref{fig:simulator} shows the A3D integration with AirSim simulator. AirSim runs on a separate simulation platform. The simulator wrapper is responsible for calling the AirSim API, forwarding captured frames and flight commands, recording experimental data, and implementing manual control of the simulator. The drone board is connected to the simulation platform via an Ethernet connection with negligible transmission delay to simulate the connection between the onboard computing device and the real drone. WiFi connection is used between the drone board and edge server for wireless communication.
Bandwidth measurements are implemented by \textit{psutil}\cite{psutil} and \textit{iperf3}\cite{iperf3}. 
All modules in A3D communicate using \textit{ZeroMQ} \cite{zeromq}. 

We use a scenario called ``Coastline'' in AirSim, which contains an approximately 1200m road with 16 turns and its typical scenes are shown in Fig. \ref{fig:coastline_scene}.
We use a Jetson Nano as the onboard computing device and a workstation with an 8-core 3.7GHz Intel CPU and 16G RAM as the edge server. 
To align with the GPU-free platform targeted in DroNet's design \cite{loquercio2018dronet}, only the CPU processor is used in evaluation, emulating the status of resource-constrained edge-assisted drones.
Additionally, we manually adjust the drone-server bandwidth based on HSDPA \cite{riiser2013commute}, a dataset that collects realistic bandwidth measurements on mobile devices, to simulate drones' wireless network fluctuations\footnote{
Wireless signal collisions in multi-drone serving are assumed to be well managed and addressed by underlying communication protocols, and have been accommodated in the bandwidth traces. In real-world deployment, one can exploit existing techniques on channel orchestration (e.g., \cite{chen2019uav, wang2018bandwidth}) to enhance A3D for addressing potential wireless signal collision issues.
}.

\section{Evaluation}
\label{sec:evaluation}
\subsection{Experimental Setup}
\label{sec:experimental_setup}
\textbf{Metric.}
Our evaluation is carried out in both the proof-of-concept prototype and simulator experiments, in order to thoroughly examine the performance of A3D.
In particular, we mainly focus on the following metrics to investigate A3D's design and optimization.
1) \textit{Quality of Navigation (QoN)}. We take the predictions of the navigation model corresponding to the configuration of zero end-to-end latency, the highest resolution ($448\times448$), and the basic image compression ratio (95\%) images as the ground truth, and use Eq. (\ref{eq:QoN}) to calculate the QoN. The ground truth represents the best performance that the employed navigation model can achieve in the most ideal case, so the measured QoN reflects the performance gap between the actual execution and the ideal case.
2) \textit{Flight distance}, a widely-used performance indicator of drone autonomy that refers to the total distance flown by the drone from the location it takes off to the location it safely lands or deviates from its course.
We repeat the flight five times to average the recorded distance.
3) \textit{End-to-end latency}, the elapsed time from the image capture to the flight command determination. Although the end-to-end latency is not our direct optimization objective, it has a significant impact on our targeted QoN performance.

\textbf{Parameters.}
The prediction error threshold $\varepsilon$ for calculating QoN is set to 1 for the prototype and 0.13 for the simulation. The time window size $\tau$ is fixed at 5s, which is equal to the length of a DRL step. For the hyper-parameters in the EIE module, $\alpha$ and $\beta$ are set to 0.3 and 0.09 respectively. When training the DRL neural scheduler, we set the length of an episode to 100s, the initial learning rate to $7\times10^{-4}$, and the discount factor $\gamma$ to 0.99. $h$ and $l$ are set 4 and 0.8, respectively.

\begin{figure}[t] 
    \centering
    \setlength{\abovecaptionskip}{-0.1cm}
    \includegraphics[width=0.8\linewidth]{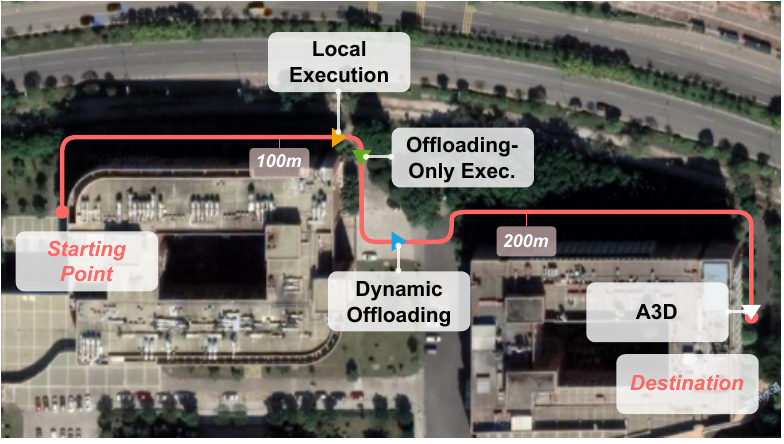}
    \caption{The campus route and the termination location of different approaches in our prototype evaluation. A3D successfully passes the complete route and safely reaches the destination.}
    \label{fig:prototype_map}
    \vspace{-0.5cm}
\end{figure}

\begin{figure*}[t] 
    \centering
    \setlength{\abovecaptionskip}{0cm}
    \subfigure[The used bandwidth traces in our experiments.]{
        \begin{minipage}[t]{0.22\textwidth}
            \centering
            \setlength{\abovecaptionskip}{-0.1cm}
            \includegraphics[height=3.1cm]{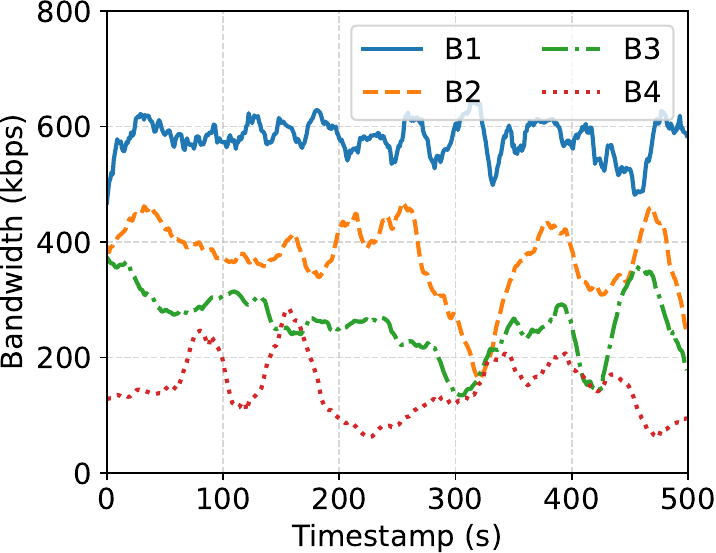}
            \label{fig:bandwidth_trace}
            \vspace{-0.5cm}
        \end{minipage}
    }
    \
    \subfigure[Quality of navigation with varying bandwidth.]{
        \begin{minipage}[t]{0.22\textwidth}
            \centering
            \setlength{\abovecaptionskip}{-0.1cm}
            \includegraphics[height=3.1cm]{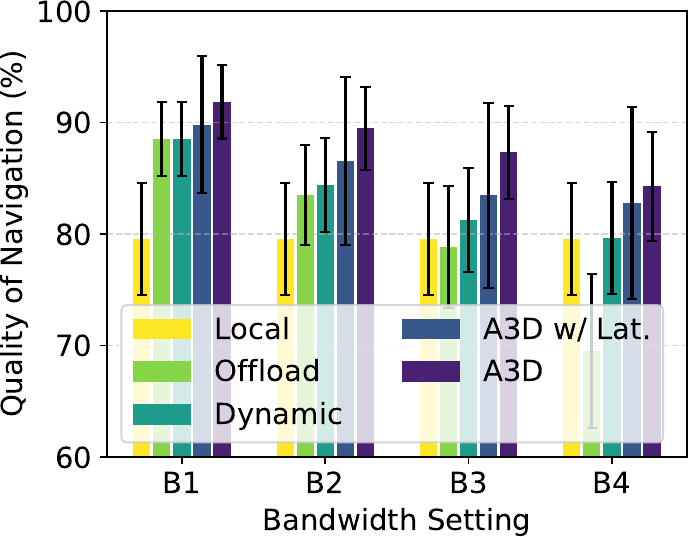}
            \label{fig:compare_acc_bandwidth}
            \vspace{-0.5cm}
        \end{minipage}
    }
    \
    \subfigure[End-to-end latency with varying bandwidth.]{
        \begin{minipage}[t]{0.22\textwidth}
            \centering
            \setlength{\abovecaptionskip}{-0.1cm}
            \includegraphics[height=3.1cm]{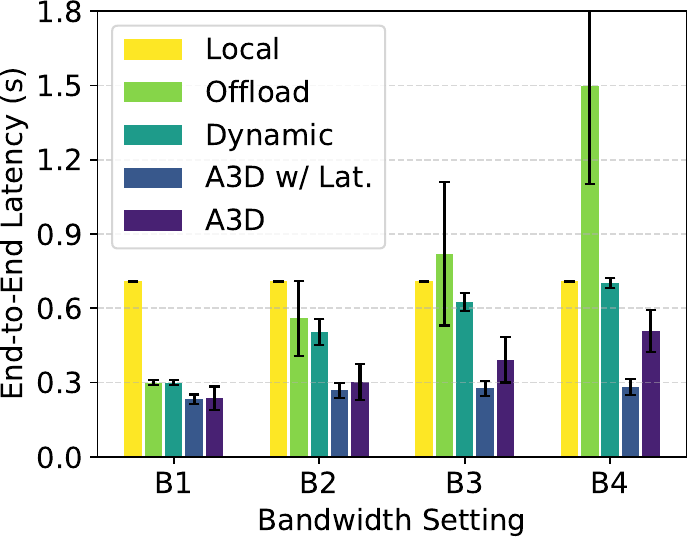}
            \label{fig:compare_latency_bandwidth}
            \vspace{-0.5cm}
        \end{minipage}
    }
    \
    \subfigure[Achieved flight distance with varying bandwidth.]{
        \begin{minipage}[t]{0.22\textwidth}
            \centering
            \setlength{\abovecaptionskip}{-0.1cm}
            \includegraphics[height=3.1cm]{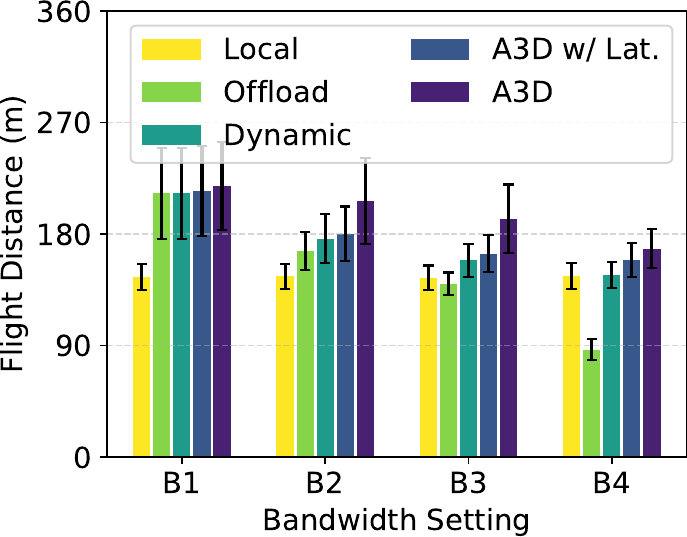}
            \label{fig:compare_acc_speed}
            \vspace{-0.5cm}
        \end{minipage}
    }
    \\
    \subfigure[Quality of Navigation with varying drone speed.]{
        \begin{minipage}[t]{0.22\textwidth}
            \centering
            \setlength{\abovecaptionskip}{-0.1cm}
            \includegraphics[height=3.1cm]{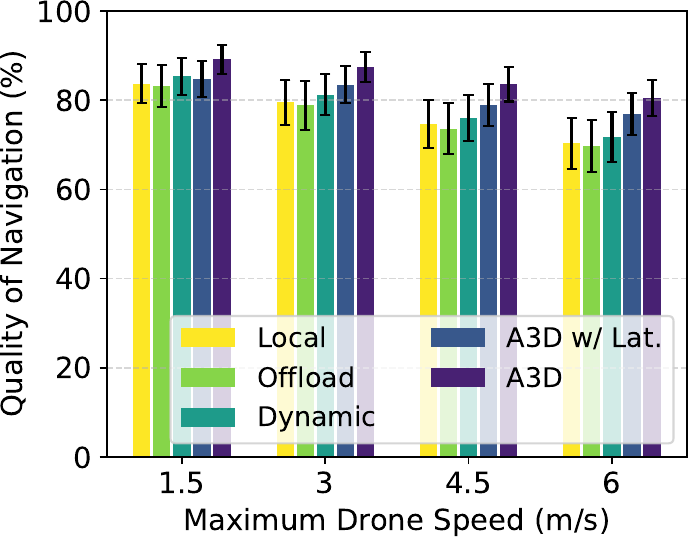}
            \label{fig:compare_qon_speed}
            \vspace{-0.5cm}
        \end{minipage}
    }
    \
    \subfigure[Achieved flight distance with varying drone speed.]{
        \begin{minipage}[t]{0.22\textwidth}
            \centering
            \setlength{\abovecaptionskip}{-0.1cm}
            \includegraphics[height=3.1cm]{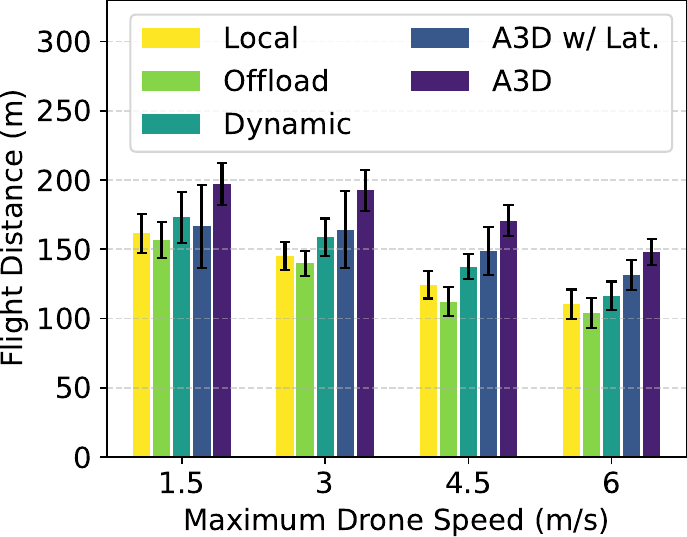}
            \label{fig:compare_distance_speed}
            \vspace{-0.5cm}
        \end{minipage}
    }
    \
    \subfigure[CDF of end-to-end latency within the flight period.]{
        \begin{minipage}[t]{0.22\textwidth}
            \centering
            \setlength{\abovecaptionskip}{-0.1cm}
            \includegraphics[height=3.1cm]{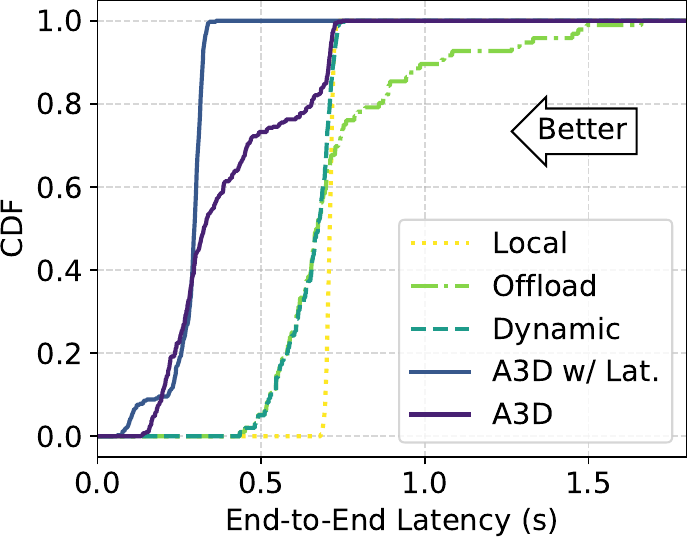}
            \label{fig:latency_cdf}
            \vspace{-0.5cm}
        \end{minipage}
    }
    \
    \subfigure[CDF of navigation model prediction error within the flight period.]{
        \begin{minipage}[t]{0.22\textwidth}
            \centering
            \setlength{\abovecaptionskip}{-0.1cm}
            \includegraphics[height=3.1cm]{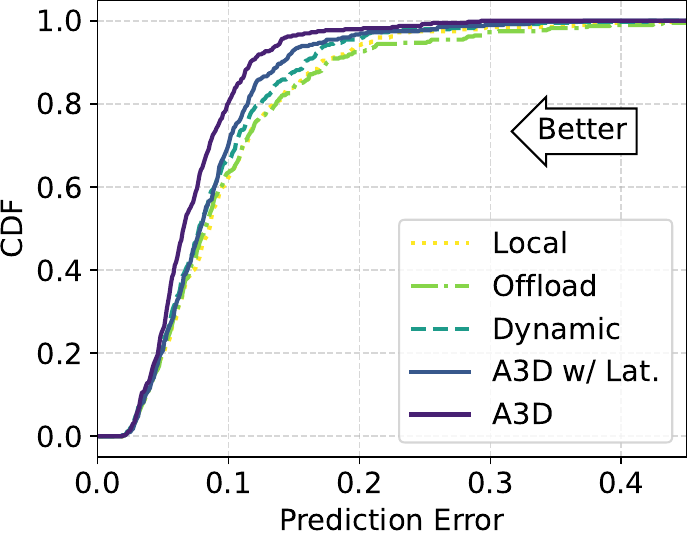}
            \label{fig:error_cdf}
            \vspace{-0.5cm}
        \end{minipage}
    }
    \caption{Single-drone evaluation results.  }
    \label{fig:single_drone_experiments}
    \vspace{-0.5cm}
\end{figure*}

\textbf{Baseline.}
We design commonly-applied heuristics as baseline strategies for single-drone and multi-drone navigation, respectively.
For single-drone evaluation, the baselines include:
1) \textit{Local}, which is a non-adaptive approach that places the navigation model on the onboard computing device for execution at any moment, using a fixed resolution ($448\times448$). This is the most common approach when the drone can carry a computing device with sufficient computation capability.
2) \textit{Offload}, which is also a non-adaptive approach that places the navigation model on the server for execution at any moment, using a fixed resolution ($448\times448$) and a fixed image compression ratio (95\%). This is a common approach when the drone has insufficient computation resources and can communicate with the server via a stable network connection.
3) \textit{Dynamic Offload (Dynamic)}. We collect experimental data to estimate the computing latency at the local or the edge, and decide the execution place by directly optimizing the end-to-end latency.
This approach merely optimizes the latency dimension by adapting the inference execution location configuration but still uses a fixed resolution and compression ratio. 

For multi-drone evaluation, the baselines are:
1) \textit{Contention-Agnostic (Agnostic)}, where drones are unaware of the existence of each other and their neural schedulers always accept the whole amount $\lambda$ as the obtained edge resources $s_t$, i.e., each drone ``believes'' that it completely possesses the whole edge resource pool.
However, the edge server will keep monitoring the connected drones at every moment and evenly allocate CPU cores for them.
2) \textit{Even}, which consistently assigns edge resources in equal proportion to every connected drone, and the drones are informed of such an even allocation results.
3) \textit{w/o Bounds}, an ablated version of A3D's resource allocation algorithm that only runs the initialization phase to generate an allocation solution.
\vspace{-0.2cm}

\subsection{Prototype Verification}

This subsection presents our experimental results on our proof-of-concept prototype in a campus route (Fig. \ref{fig:prototype_route}).
Fig. \ref{fig:prototype_distance_latency} depicts the complexity of this route and the measured inference latency of the navigation model when executing at the drone board locally and the edge server.
For each flight tour, we set the inference latency as a determined value and let the drone fly freely until it turns off track, following the same methodology in Fig. \ref{fig:motivation_latency}'s setting.
From the figure, we observe that the accomplished flight distance dramatically diminishes as the navigation decision latency increases.
If the drone computes the navigation decisions by itself, it flies around 140m, while a pure offloading solution attains a similar meterage.
In particular, if the end-to-end latency reaches 0.9s, the drone yaws at the beginning, implying that it fails to pass the first bend at the starting point.

We next investigate the distribution of navigation model prediction accuracy and end-to-end latency within the flight period and plot the results in Fig. \ref{fig:prototype_accuracy_distribution} and Fig. \ref{fig:prototype_latency_distribution}.
\textit{Local} and \textit{Offload} hold much more significant prediction errors due to the high end-to-end latency. \textit{Dynamic} method decreases the prediction error by simple optimization while \textit{A3D} retains the lowest prediction error by comprehensive optimization towards QoN.
As for the latency, the real-life experiment results maintain strong consistency with that in simulation (Sec. \ref{sec:single_drone_experiment}). The latency of \textit{Local} is distributed around 588ms because its computing only relies on the onboard processor. \textit{Offload} is highly affected by the wireless drone-edge connection and its latency measurements has the most significant variance. \textit{Dynamic} switches its execution location concerning the latency and approximately records the lower bound of \textit{Local} and \textit{Offload}. 
By contrary, A3D holds the lowest end-to-end latency owing to its ability of jointly adjusting configurations in the design space of scheduling adaptability. 

Fig. \ref{fig:prototype_qon} displays A3D's achieved QoN at different maximum flight speeds against baselines.
We set the prediction error threshold $\varepsilon$ as 1 to maximize the expressiveness in the real-life environment.
The figure shows that A3D clearly obtains the highest QoN among other approaches across different maximum speeds.
Specifically, A3D improves the QoN by up to 21.97\% compared to \textit{Local}.
The faster the speed is, the more performance improvement the A3D gains.
This is because higher flight speed introduces faster scenarios transition, emphasizing the necessity of lower end-to-end latency. 
The QoN of A3D shows little changes with various maximum flight speeds since A3D's adaptive configuration can significantly mitigate the latency issue, demonstrating its practicability.

Fig. \ref{fig:prototype_map} visualizes the termination locations of the four approaches. \textit{Local} yaws to the right too late because of the high inference latency on the device and fails to pass at the second 90-degree bend. \textit{Offload} holds a similar flight distance as \textit{Local} and it also cannot pick a proper moment to turn around. \textit{Dynamic} succeeds to conquer the second turn owing to its adaptability to choosing the execution location.
However, this method is empirical and environment-agnostic, resulting in the yaw when meeting consecutive bends.
A3D keeps the superior performance and manages to fly the complete route while the others fail halfway.
This attributes to A3D's neural scheduler that can adaptively adjust system configurations to strike a balance in the latency-accuracy tradeoff.
\vspace{-0.2cm}

\begin{figure*}[t] 
    \centering
    \setlength{\abovecaptionskip}{0cm}
    \subfigure[Average Quality of Navigation with varying edge resources.]{
        \begin{minipage}[t]{0.22\textwidth}
            \centering
            \setlength{\abovecaptionskip}{-0.1cm}
            \includegraphics[height=3.1cm]{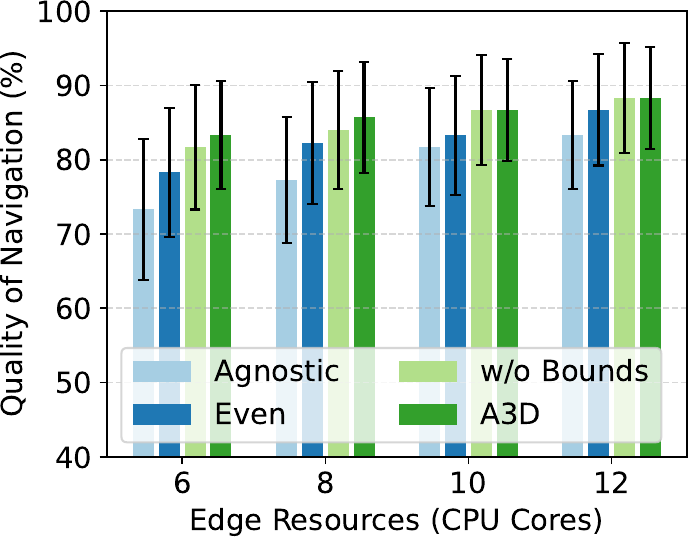}
            \label{fig:qon_multi_drone}
            \vspace{-0.5cm}
        \end{minipage}
    }
    \
    \subfigure[Average flight distance with varying edge resources.]{
        \begin{minipage}[t]{0.22\textwidth}
            \centering
            \setlength{\abovecaptionskip}{-0.1cm}
            \includegraphics[height=3.1cm]{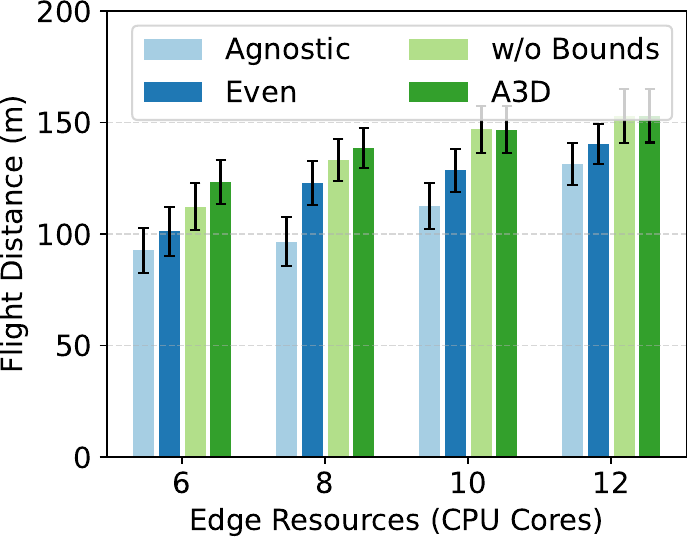}
            \label{fig:distance_multi_drone}
            \vspace{-0.5cm}
        \end{minipage}
    }
    \
    \subfigure[Average end-to-end latency with varying edge resources.]{
        \begin{minipage}[t]{0.22\textwidth}
            \centering
            \setlength{\abovecaptionskip}{-0.1cm}
            \includegraphics[height=3.1cm]{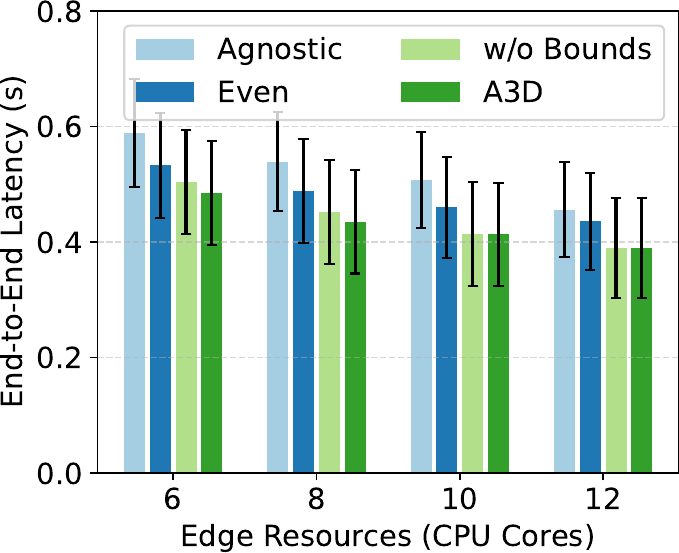}
            \label{fig:latency_multi_drone}
            \vspace{-0.5cm}
        \end{minipage}
    }
    \
    \subfigure[Average offloading ratio with varying edge resources.]{
        \begin{minipage}[t]{0.22\textwidth}
            \centering
            \setlength{\abovecaptionskip}{-0.1cm}
            \includegraphics[height=3.1cm]{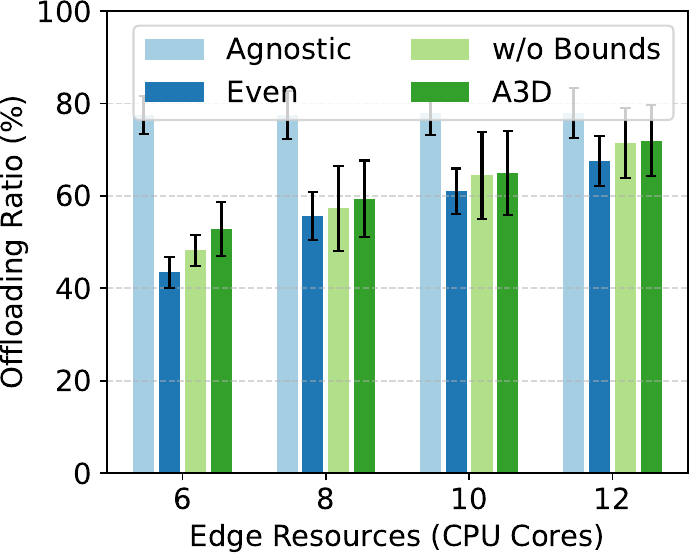}
            \label{fig:offloading_ratio_multi_drone}
            \vspace{-0.5cm}
        \end{minipage}
    }
    \\
    \subfigure[CDF of prediction errors for all drones.]{
        \begin{minipage}[t]{0.22\textwidth}
            \centering
            \setlength{\abovecaptionskip}{-0.1cm}
            \includegraphics[height=3.1cm]{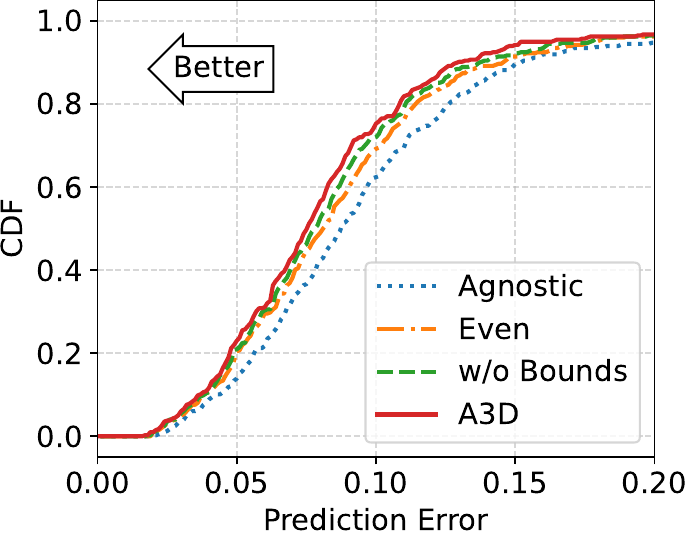}
            \label{fig:qon_error_multi_drone}
            \vspace{-0.5cm}
        \end{minipage}
    }
    \
    \subfigure[CDF of end-to-end latency for all drones.]{
        \begin{minipage}[t]{0.22\textwidth}
            \centering
            \setlength{\abovecaptionskip}{-0.1cm}
            \includegraphics[height=3.1cm]{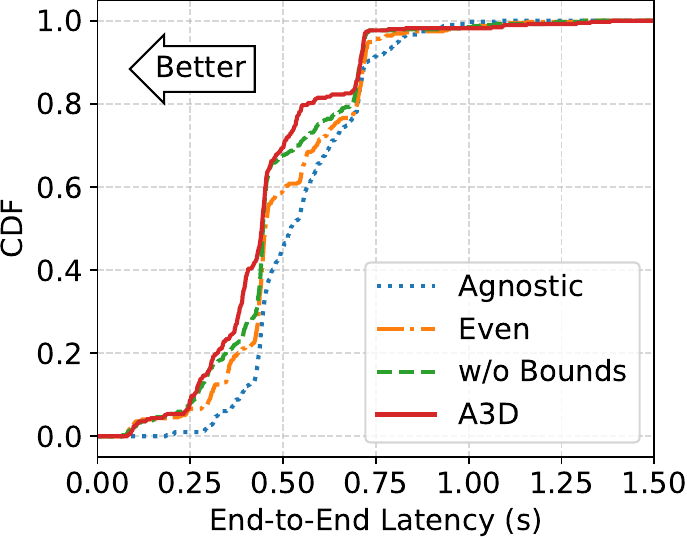}
            \label{fig:cdf_multi_drone}
            \vspace{-0.5cm}
        \end{minipage}
    }
    \
    \subfigure[Average Quality of Navigation with varying number of drones.]{
        \begin{minipage}[t]{0.22\textwidth}
            \centering
            \setlength{\abovecaptionskip}{-0.1cm}
            \includegraphics[height=3.1cm]{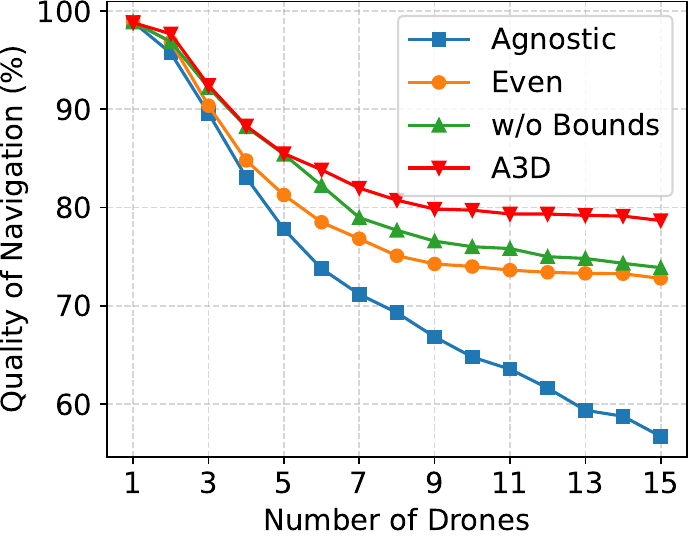}
            \label{fig:qon_num_drone}
            \vspace{-0.5cm}
        \end{minipage}
    }
    \
    \subfigure[Average offloading ratio with varying number of drones.]{
        \begin{minipage}[t]{0.22\textwidth}
            \centering
            \setlength{\abovecaptionskip}{-0.1cm}
            \includegraphics[height=3.1cm]{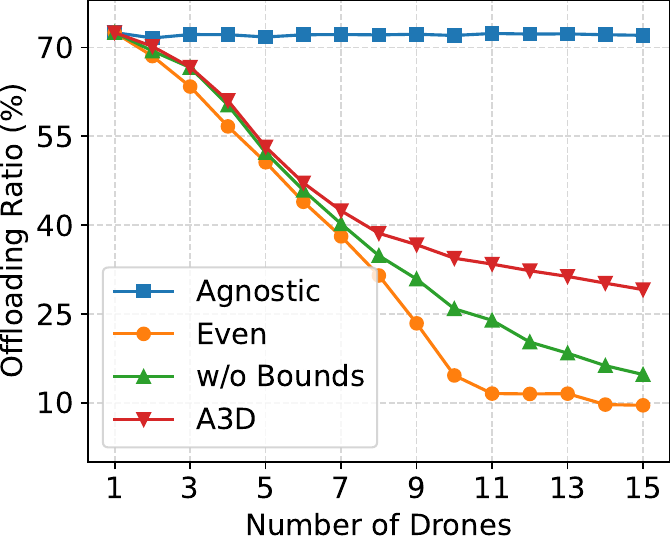}
            \label{fig:ratio_num_drone}
            \vspace{-0.5cm}
        \end{minipage}
    }
    \caption{Multi-drone evaluation results.  }
    \label{fig:multi_drone_experiments}
    \vspace{-0.5cm}
\end{figure*}

\subsection{Performance Comparison with Single Drone}
\label{sec:single_drone_experiment}
This subsection evaluates A3D in our simulation testbed under single-drone settings.
To demonstrate the effectiveness of our proposed QoN metric, we further compare an ablated version of A3D, marked as \textit{A3D w/ Lat.}, by training the DRL scheduler using latency as the reward.
First, we assess the performance of A3D in different bandwidth conditions. We pick four bandwidth traces in the dataset \cite{riiser2013commute}, which collect real-world traces and are labeled in Fig. \ref{fig:bandwidth_trace} as B1, B2, B3, and B4, respectively. As shown in Fig. \ref{fig:compare_acc_bandwidth}, A3D achieves the highest QoN across all bandwidth conditions. When the bandwidth decreases, \textit{Offload}'s QoN decreases significantly, which is caused by the rise in transmission delay. In contrast, \textit{Local} is independent of bandwidth as it isolates drones from edge servers.
\textit{Dynamic}'s QoN is always slightly higher than the local and offload baselines, suggesting that dynamically choosing whether to offload or not can improve performance.
However, since \textit{Dynamic} does not adjust its choice of resolution and image compression ratio, it fails to reach the same performance improvement as A3D. Fig. \ref{fig:compare_latency_bandwidth} shows the end-to-end latency of these methods, where \textit{A3D w/ Lat.}'s results are always the lowest since it directly optimizes latency as the scheduling objective.
As for the original A3D trained with QoN, its latency is reduced by 28.06\% compared with \textit{Dynamic} at B4, indicating that A3D can intelligently and jointly adjust the offloading decision, image resolution, and compression ratio so as to strike a better balance between accuracy and latency.

We next evaluate the performance of A3D at different flight speeds. We set the maximum drone speed $v_{max}$ to 1.5m/s, 3m/s, 4.5m/s, and 6m/s respectively, and the results are shown in Fig. \ref{fig:compare_qon_speed} and Fig. \ref{fig:compare_distance_speed}. A3D achieves a higher QoN of navigation than baselines at all speeds, and is able to improve the QoN by 4\%-12\%.
The faster the speed, the greater the A3D's improvement gains.
Fig. \ref{fig:compare_distance_speed} shows the results on flight distance. Specifically, A3D is able to achieve a 5.68\%-27.28\% improvement, which is greater than the QoN improvement shown in Fig. \ref{fig:compare_acc_speed}.
The reason is that the drone is less fault-tolerant at higher speeds and a few prediction errors can cause the drone to deviate from its course, meaning the principle of minimizing prediction errors in A3D can validly improve flight distance. Both the two figures indicate a tight correlation between flight distance and QoN, demonstrating that using QoN as the reward can provably improve drones' flying ability.

We further investigate the distribution of each metric. Fig. \ref{fig:latency_cdf} shows the Cumulative Distribution Function (CDF) of end-to-end latency for the B3 trace in Fig. \ref{fig:compare_acc_bandwidth}. The latency of \textit{Local} is distributed around 700ms since it only uses the dedicated onboard resource. \textit{Offload}'s latency rises significantly when the bandwidth is low and thus a proportion of its distribution lies at a higher level ($>$750ms).
\textit{Dynamic}'s result is the lower bound of \textit{Local}'s and \textit{Offload}'s, but it is still much higher than A3D's because A3D can reduce the latency by adjusting images' resolution and compression ratio. 
Fig. \ref{fig:error_cdf} shows the CDF of the navigation model's prediction errors at the steering angle for the B3 trace.
A3D can achieve lower prediction errors than baselines, consistent with the results above. 
Interestingly, while A3D falls short in latency performance compared with \textit{A3D w/ Lat.}, it achieves better prediction performance in Fig. \ref{fig:error_cdf}, which implies the optimization tradeoff implicated in the QoN metric.

\vspace{-0.2cm}

\subsection{Performance Comparison with Multiple Drones}

This subsection examines A3D's resource allocation algorithm in our simulation testbed under multi-drone settings.
Specifically, we use four Jetson Nanos to emulate four drones, and accordingly launch four UAV instances in AirSim.
Their maximum flight speed is fixed at 3m/s, and their networking conditions towards the edge server follow the bandwidth traces B1, B2, B3, and B4 in Fig. \ref{fig:bandwidth_trace}, respectively.
An experiment trial is finished when one of the drones yaws on the route, and their average performance measurements are recorded as the results.

Fig. \ref{fig:qon_multi_drone}-Fig. \ref{fig:offloading_ratio_multi_drone} displays A3D's performance in different dimensions: QoN, flight distance, end-to-end latency, and offloading ratio.
In particular, Fig. \ref{fig:qon_multi_drone} and Fig. \ref{fig:distance_multi_drone} show that A3D always yields the highest QoN and flight distance over other counterparts, achieving up to 13.6\% QoN improvement and extending the average flight distance of drones for at most 42.07m.
In contrast, the \textit{Agnostic} approach records a poor performance across setups, and the gap between it and A3D widens when assigned CPU cores are fewer.
This reveals the necessity of the resource allocator module, especially when edge resources are limited.
\textit{Even} approach performs better than \textit{Agnostic}, but still falls short compared to A3D and its ablated version (\textit{w/o Bounds}).
The difference between A3D with and without bounds is small when edge resources are abundant ($geq$10 CPU cores).
This is because with more edge resources the initialized allocation usually has satisfied the requirement of a bounded interval, and does not need the bounded reallocation phase anymore.
Conversely, in an edge server with limited edge resources, the bounded reallocation can effectively align resource allocation to avoid resource waste and thus boost global performance. The end-to-end latency results in Fig. \ref{fig:latency_multi_drone} exhibit a strong correlation with results in Fig. \ref{fig:qon_multi_drone}, where A3D continuously attains the lowest latency within the flight.
This also reflects better resource utilization of A3D against other baselines.
Fig. \ref{fig:offloading_ratio_multi_drone} plots the offloading ratios of different approaches during the flight, which calculates the percentage of offloading period out of the total flight period.
For \textit{Agnostic} approach, the offloading ratio logs in a high level because its perceived edge resource is always the amount of the resources in the edge server.
However, it does not translate frequent offloading into a high QoN in Fig. \ref{fig:qon_multi_drone}, because the actual resources that the drone can utilize are inconsistent with what they see and are impacted by potential contention.
For other approaches, their comparison on offloading ratio appears in a similar pattern to that in the QoN dimension, where A3D with the highest offloading ratio witnesses the highest QoN.
By judiciously allocating resources for drones, A3D can encourage the drones to utilize the edge server's assist and consequently promote the overall system performance.

Fig. \ref{fig:qon_error_multi_drone} and Fig. \ref{fig:cdf_multi_drone} respectively depict the CDF of the prediction errors and the end-to-end latency of all drones during the whole flight.
In Fig. \ref{fig:qon_error_multi_drone}, we observe that the four approaches have close trajectories of prediction errors, while in Fig. \ref{fig:cdf_multi_drone} A3D's latency distribution is clearly lower.
This validates the results in Fig. \ref{fig:qon_multi_drone} and Fig. \ref{fig:latency_multi_drone}, where A3D outperforms other baselines for all cases.

To further investigate the performance of A3D's resource allocation with more drones, we carry out numerical simulations using the data traces collected from real drones.
We fix an amount of edge resources at 12 CPU cores and vary the number of drones from 1 to 15.
Fig. \ref{fig:qon_num_drone} and Fig. \ref{fig:ratio_num_drone} give the QoN and offloading ratio results, respectively.
For \textit{Agnostic}, its resource information blindness implies the inconsistency between how much resource drones require and how much resource edge servers provide, and can thus result in resource contention at the edge server.
As the number of drones grows, the resource contention becomes increasingly intensive and therefore \textit{Agnostic}'s QoN drops quickly.
\textit{Even} approach enforces all drones to share equal opportunities to edge resources and allows them to see how much they will obtain.
Under this mechanism, each drone's obtained resources shrink with the system connecting more drones, which reduces the possibilities of their offloading decision (as indicated in Fig. \ref{fig:ratio_num_drone}), wastes edge resources, and thereupon lower their achieved QoN.
In contrast to \textit{Agnostic} and \textit{Even}, \textit{w/o Bounds} can estimate the demand of each drone based on their server connectivity, and accordingly assign edge resources in an on-demand manner.
However, without the bounded reallocation phase in A3D’s algorithm, this approach may still lead to inefficient resource utilization since the benefit of resource supplement diminishes marginally as illustrated in Fig. 13(a).
In Fig. \ref{fig:qon_num_drone}, though its performance is on par with A3D when the number of drones is small ($<$6), its QoN results go closer to \textit{Even} when the number of drones grows.
By contraries, A3D employs a bounded reallocation to drop a part of services to ensure the QoN of remaining drones, which yet achieves better global system performance.
Fig. \ref{fig:ratio_num_drone} shows the offloading ratio with varying number of drones.
As the drones in \textit{Agnostic} are only aware of a constant edge resource $\lambda$, its offloading ratio results is independent of the number of drones.
For \textit{Even} and \textit{w/o Bounds}, their offloading ratio quickly descends, implying a tendency of using on-board computing resources.
A3D's offloading ratio is higher than \textit{Even} and \textit{w/o Bounds}, which demonstrates a better resource efficiency and confirms the superior QoN in Fig. \ref{fig:qon_num_drone} over other counterparts.
\vspace{-0.2cm}

\begin{figure}[t]
\centering
\setlength{\abovecaptionskip}{-0.1cm}
\includegraphics[width=0.8\linewidth]{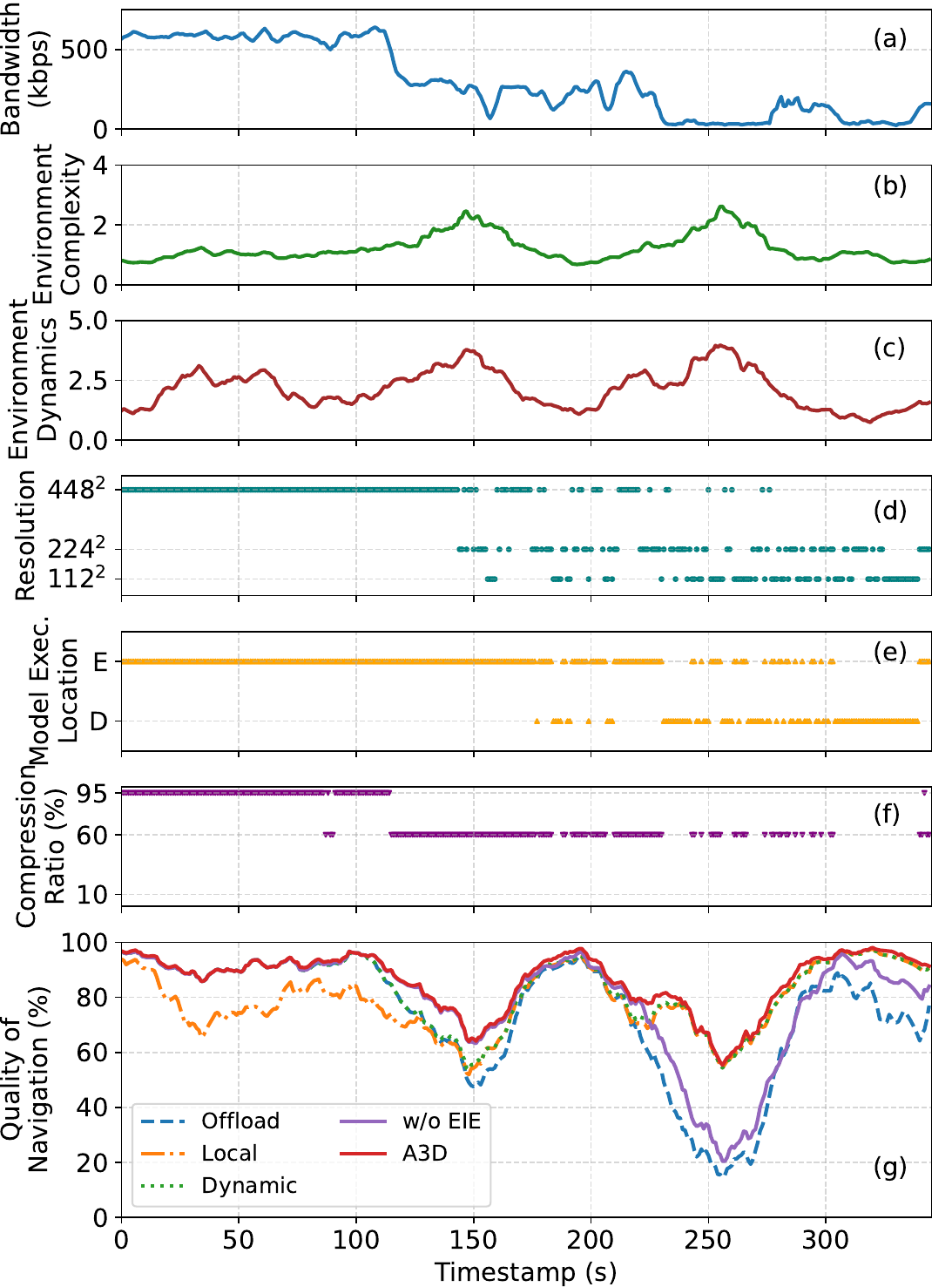}
\caption{
Case study of A3D's adaptive decisions. (a) Bandwidth trace. (b) Environment complexity records. (c) Environment dynamics records. (d) A3D's input resolution decisions. (e) Decided execution locations of navigation model inference, where D and E stand for onboard device and server, respectively. (f) A3D's compression ratio decisions. (g) Quality of Navigation of different approaches during the flight period, where EIE indicates the neural scheduler's environmental information encoding module.
}
\label{fig:case_study}
\vspace{-0.6cm}
\end{figure}

\subsection{Adaptability}
\label{sec:adaptability}
This subsection investigates how A3D makes dynamic decisions to adapt to the environment.
Using A3D (with its ablated version) and three baselines, we perform a flight of 350s long in AirSim simulator with a maximum flight speed limited to 3m/s. When the drone deviates from its course, we manually control the drone to return to the correct direction.
Fig. \ref{fig:case_study}(a) shows the bandwidth trajectory of the entire flight. Fig. \ref{fig:case_study}(b) and (c) show the fluctuation of Environment Complexity and Dynamics (defined in Eq. (\ref{eq:env_complexity}) and Eq. (\ref{eq:env_dynamic}), respectively). Fig. \ref{fig:case_study}(d), (e), and (f) illustrate the selection of three decision variables of A3D, i.e., input resolution, inference execution location, and compression ratio. During the first 120 seconds, A3D always chooses to offload the model to the server because the bandwidth is at a high level, and offloading to the server will provide more benefits. According to Fig. \ref{fig:case_study}(a), there is a significant decrease in bandwidth around 120 seconds, to which A3D responds by reducing the compression ratio from 95 to 60 to reduce the amount of data transferred. In the middle to late stages of the experiment, as bandwidth remains low, A3D begins to alternate between local computation and offloading to the server: the lower the bandwidth, the more likely A3D will choose to compute locally. For the choice of resolution, A3D gradually switches from the highest resolution to a lower resolution for inference to reduce latency. Considering the high computing latency brought by a high-resolution input, A3D prefers to resize an image in a lower resolution when computing the navigation model locally. The above results validate that A3D always achieves higher QoN compared to the other baselines, shown in Fig. \ref{fig:case_study}(g). 
We also inspect the effectiveness of the Environmental Information Encoding (EIE) module by deactivating it in A3D's neural scheduler.
As the bandwidth declines and the environment becomes more complex and highly dynamic (timestamp [200,300]), however, \textit{A3D w/o EIE} significantly drops its QoN and performs even worse than \textit{Local} baseline.
This implies that the system without EIE can still possess the ability of adaptive scheduling, which however is relatively limited compared to the complete A3D (with EIE).
Such mild adaptability comes from the capability of DRL's neural network agent, but the lack of EIE makes it fall short in environments with extremely low bandwidth and complex scenes, which is exactly what EIE-enhanced A3D can deal with.
\vspace{-0.3cm}

\begin{figure}[t]
\centering
    \setlength{\abovecaptionskip}{-0.1cm}
    \includegraphics[height=3cm]{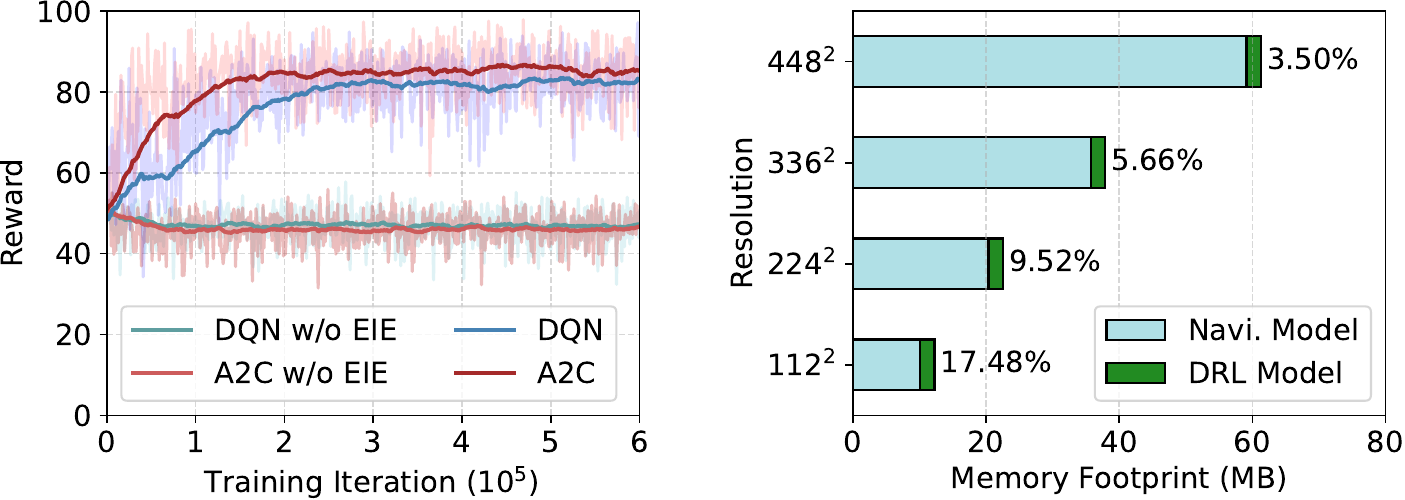}
    \caption{
    Left: training curves of the neural scheduler with different DRL models with and without the Environmental Information Encoding (EIE) module.
    Right: The memory footprint of the neural scheduler and the navigation model.
    }
    \label{fig:neural_scheduler}
    \vspace{-0.5cm}
\end{figure}

\begin{table}[t]
\caption{Comparison of varying scheduler configurations.}
\label{table:drl_configuration}
\centering
\begin{tabular}{|c|c|c|c|}
    \hline
    \begin{tabular}[c]{@{}c@{}}\textbf{Actor}\\\textbf{Network}\end{tabular}     &  \begin{tabular}[c]{@{}c@{}}\textbf{Critic}\\\textbf{Network}\end{tabular}       & \begin{tabular}[c]{@{}c@{}}\textbf{Convergent}\\\textbf{Reward}\end{tabular} & \begin{tabular}[c]{@{}c@{}}\textbf{Execution}\\\textbf{Overhead (ms)}\end{tabular} \\ \hline \hline
    {[}64,64{]}       & {[}64,64{]}   & 84.65$\pm$4.618                                                 & 4.35                                                              \\ 
    {[}128,128{]}     & {[}128,128{]} & 84.70$\pm$4.501                                                 & 4.67                                                              \\ 
    {[}128,128{]}     & {[}32,32{]}   & 84.63$\pm$4.957                                                 & 4.67                                                              \\ 
    \textbf{{[}128,128{]}}     & \textbf{{[}64,64{]}}   & \textbf{85.08}$\pm$\textbf{4.178}                                                 & \textbf{4.67}                                                              \\ 
    {[}128,128,128{]} & {[}64,64{]}   & 83.49$\pm$7.898                                                 & 6.35                                                              \\ 
    {[}256,256{]}     & {[}64,64{]}   & 84.42$\pm$5.581                                                 & 5.53                                                              \\ \hline
    \end{tabular}
    \vspace{-0.5cm}
\end{table}

\subsection{Neural Scheduler Implication}
\label{sec:eval_scheduler}
Our neural scheduler is implemented using the \textit{stable-baseline} framework\cite{stable-baselines3} based on Pytorch, with RMSProp adopted as the optimizer. We explore the optimal structure of Actor and Critic networks in the DRL model.
To find the best parameterized configuration, we use different network structures and calculate the mean and variance of their rewards after convergence, as listed in Table \ref{table:drl_configuration}, where bracketed values represent the number of neurons in the hidden layer.
The experiments show that the DRL model converges with the highest rewards and the lowest variance with a two-layer 128-neuron Actor network and a two-layer 64-neuron Critic network, which is therefore set as the default structure through other evaluations.
Fig. \ref{fig:neural_scheduler}(left) shows the training curves of the DRL model's reward (QoN) with the total $6\times 10^{5}$ training iterations, which takes about 8 hours in the numerical simulation environment.
we compare two DRL algorithms, A2C and Deep Q-Network (DQN), and it can be seen that A2C's both convergence speed and convergent reward are better than DQN. 
We also witness that without the EIE module, both algorithms' rewards fail to climb to a higher altitude given thousands of training iterations.
Their curves remain at a much lower level than that of the original version (A2C/DQN with EIE), implying that the absence of EIE could lead to invalid optimization towards QoN and confirms the limited scheduling adaptability in the case study experiment (Sec. \ref{sec:adaptability}).

We also examine the overhead of our neural scheduler. We measure the execution overhead on the onboard device (Jetson Nano) and the results are shown in Table \ref{table:drl_configuration}. It can be seen that the execution overhead is around 5ms for all network structures, which is negligible in the whole framework.
In addition, we compare the memory overhead of the DRL model and the navigation model in Fig. \ref{fig:neural_scheduler}(right). The memory footprint of the navigation model is tightly related to the resolution of input images. Specifically, the memory space taken by the DRL model is 3.50\%-17.48\% out of the whole. For any resolution, the memory footprint of the DRL scheduler is much lower than that of the navigation model, indicating that it is minority compared to the core of navigation tasks.
\vspace{-0.2cm}

\section{Related Work}
\label{sec:related_work}
\textbf{Autonomous drone navigation.}
With the successful application of CNN in computer vision, more and more research has used CNN for drone navigation and obstacle avoidance.
In \cite{gandhi2017learning}, a self-supervised learning approach is used to train an image classification CNN to achieve autonomous drone obstacle avoidance indoors. 
The authors in \cite{smolyanskiy2017toward} use their dataset collected on foot to train an image regression CNN model to predict the drone's turn angle and achieve autonomous drone navigation along a forest trail. \cite{kang2019generalization} trains a navigation model for predicting turn angles in a drone simulator to achieve autonomous drone navigation and obstacle avoidance indoors. 
These works use CNN to directly control drones, ignoring the decisions that A3D optimizes.

\textbf{Edge computing for drones.}
Drones, as end devices that often perform computationally intensive tasks, can gain many benefits from edge computing\cite{chen2019uav}, especially for vision-based drone tracking\cite{EyeIntheSky} and detection\cite{greyon2022, gumaei2021deep}.
\cite{wang2018bandwidth} proposes a framework to minimize the amount of transmitted data while ensuring the accuracy of drone video analysis with edge-assisted.
\cite{chinchali2020sampling} proposes a method to reduce the amount of data transmission when robots and edge servers jointly train a model. 
The authors in \cite{hayat2021edge} and \cite{messous2020edge} both study the scenarios in which an edge server assists a drone to perform SLAM in order to reduce the latency and energy consumption of the drone. This line of research does not consider the CNN-based navigation model which requires better state abstraction modules to facilitate our DRL-based online scheduling algorithm.  

\textbf{DRL for task scheduling.} DRL is widely recognized as a promising tool to solve scheduling problems given its powerful learning capability for online decision making. 
Pensieve\cite{mao2017neural} uses DRL to automatically learn an adaptive bitrate policy to optimize various Quality of Experience (QoE) metrics.
\cite{mao2019learning} proposes a DRL-based scheduler called Decima that learns workload-specific scheduling policies for complex data processing jobs. 
For video streaming analysis, AdaDeep\cite{liu2018demand} integrates a combination of parameter pruning, matrix decomposition, and model structure replacement at different layers, using DQN to select the best compression model at runtime based on the accuracy, latency, memory, and energy requirements provided by the user.
\cite{zhao2021edgeml} proposes an edge-assisted scheduling system EdgeML that uses DRL to learn model partitioning and early exit policies to meet user requirements on latency, energy, and accuracy. 
Compared with the above works, our DRL environment has more complex state feature dependencies affecting the optimal actions and needs new design modules embedded to accommodate a CNN-based drone navigation network.  
\vspace{-0.3cm}

\section{Conclusion}
\label{sec:conclusion}
In this paper, we propose A3D, an edge-assisted cooperative drone navigation framework for high-quality autonomous flight.
By treating adaptive navigation as a service and designing a DRL-based scheduler, A3D is able to dynamically adjust the resolution, model execution position, and image encoding quality according to the changes of the environment and networking conditions. 
To support high-quality multi-drone serving, A3D develops a network-aware resource allocation algorithm to judiciously assign proper edge resources for the corresponding serving containers.
Extensive evaluation based on a proof-of-concept prototype and simulation demonstrates its effectiveness and efficiency, showing that A3D can improve 27.28\% flight distance and reduce 28.06\% latency compared to non-adaptive solutions.
\vspace{-0.3cm}

\ifCLASSOPTIONcaptionsoff
  \newpage
\fi

\bibliographystyle{IEEEtran}
\bibliography{main.bib}

\end{document}